\documentclass[doublespacing]{elsart1}
\usepackage{natbib}
\usepackage{graphics}
\usepackage{graphicx}
\usepackage{epsfig}
\usepackage{amssymb}
\usepackage{amsfonts}
\usepackage{amsmath}

\begin{document}

%\tableofcontents
\setlength{\oddsidemargin}{-0.4cm} \setlength{\evensidemargin}{-0.4cm}

\begin{frontmatter}

\title{Bouc-Wen-type models with stiffness degradation: thermodynamic analysis and applications}

\author[a]{Silvano Erlicher\corauthref{cor1}},
\ead{erlicher@lami.enpc.fr}
\author[c]{Oreste S. Bursi, A.M. ASCE}
\corauth[cor1]{Corresponding author. Tel: +33 1 64 15 37 80, Fax:
+33 1 64 15 37 41}

\address[a]{
%Laboratoire Analyse des Mat\'{e}riaux
  %et Identification LAMI
Institut Navier (LAMI/ENPC), 6 et 8 av. B. Pascal, Cit\'{e}
Descartes, Champs-sur-Marne, 77455 Marne-la-Vall\'{e}e, Cedex 2,
France}

\address[c]{Universit\`{a} di Trento, Dipartimento di Ingegneria
Meccanica e Strutturale DIMS, Via Mesiano 77, 38050, Trento, Italy}

\begin{abstract}
In this paper, a thermodynamic analysis of Bouc-Wen models endowed
with both strength and stiffness degradation is provided. It is
based on the relationship between the flow rules of these models and
those of the endochronic plasticity theory with damage, discussed in
a companion paper \citep{ErlicherPointA}. Using the theoretical
framework of that extended endochronic theory, it is shown that an
elastic Bouc-Wen model with damage, i.e. without plastic strains,
can be formulated. Moreover, a proper definition of the dissipated
energy of these Bouc-Wen models with degradation is given and some
thermodynamic constraints on the parameters defining the models
behavior are emphasized and discussed. In particular, some
properties of the energetic linear stiffness degradation rule as
well as the so-called pivot rule, well-known in the seismic
engineering field, are illustrated and commented upon. An improved
energetic stiffness degradation rule and a new stiffness degradation
rule are proposed.
\end{abstract}

\begin{keyword}

Bouc-Wen-type models \sep Stiffness degradation \sep Thermodynamics
\sep Damage \sep Dissipated energy  \sep Pivot rule \sep Energetic
linear rule.

ASCE Manuscript number: EM/2006/024438

\end{keyword}

\end{frontmatter}

\newpage

\clearpage\clearpage

\section{Introduction}

\citet{Bouc67,Bouc71} developed an univariate model for structural
and ferro-magnetic applications. The Bouc's formulation was similar
to the one proposed by \citet{Volterra28} to represent hereditary
phenomena. However, the clock-time in the Volterra-Stieltjes
integrals was substituted by an \emph{internal} or \emph{intrinsic}
\emph{time}. The Bouc model was modified by the contributions of
several authors; see, among others, \citet{Wen76}, \citet{Karray89}
and \citet{Casciati89}, leading to a more general class of models,
hereafter named Bouc-Wen-type (BW) models. A further important
modification concerned the introduction of the so-called strength
and stiffness degradation effects, by means of suitable degradation
functions \citep{Baber81}. In fact, collapse assessment in
earthquake engineering requires hysteretic models that include
strength and stiffness deterioration properties
\citep{Ibarra_et_al2005}. Some general assumptions on the form of
these functions were provided, e.g. they have to be positive and
monotone, but no other limitation was mentioned. The resulting
hysteresis models were extensively used in the seismic analysis of
structural components, e.g. \citet{Reinhorn et al 95, Foliente95}.
Moreover, the simplicity of BW models, due to the absence of an
elastic domain, allowed their use in stochastic structural analysis.
More recently, the BW models were used in applications of structural
control, in particular in the modeling of the behavior of
magneto-rheological dampers \citep{Sain97, Jansen2000} or other
kinds of damping devices \citep{Ikhouane_et_al2005}. Moreover,
several contributions of the last three decades were concerned with
the identification of BW model parameters, for applications in
seismic engineering; e.g. \citet{Masri et al 2004}.

Starting from a different viewpoint and independently from Bouc,
\citet{Valanis71} proposed the \emph{endochronic (EC) theory of
visco-plasticity}, which postulates the existence of an intrinsic
time governing the rate-independent evolution of stress and strain
in materials, whereas the Newtonian time is used to model the
viscous behavior. In the case of plasticity without viscous effects,
it was proven that flow rules can be formulated by using the
mathematical formalism of pseudo-potentials. This was done for the
standard EC theory by \citet{ErlicherPoint2005} and for the EC
theory with isotropic damage in the companion paper
\citep{ErlicherPointA}.

A major question concerning the BW models is their thermodynamic
admissibility. Several criticisms were moved to these models as
their formulation was not based on a proper thermodynamic approach.
Recently, this problem was analyzed by \citet{Ahmadi97} and
\citet{Capecchi2001} for models without degradation terms. On the
other hand, the typical constitutive laws of EC models are
characterized by the absence of an elastic domain and the
corresponding hysteresis loops are smooth and open, like in the BW
models. For this reason several authors, among others,
\citet{Casciati89} and \citet{Sivaselvan2000}, observed that a
relationship must exist between EC and BW models. A formal proof
that all the Bouc-Wen type models with a \emph{strength} degradation
term admit an equivalent endochronic formulation was provided by
\citet{ErlicherPoint2004}. This was sufficient to prove that BW
models with strength degradation are thermodynamically admissible,
in the sense that they fulfill the second principle.

Nonetheless, a problem remains unsolved, viz. the formal proof of
the thermodynamic admissibility of BW models endowed with a
\emph{stiffness} degradation term: this is the subject of the
present paper. In this respect, the endochronic theory with damage
discussed in \citet{ErlicherPointA} is used: it is proved that for
every BW model with strength and stiffness degradation, an
equivalent EC model with damage can be defined, i.e. a model
exhibiting the same flow rules. Therefore, the BW model is
thermodynamically admissible if the corresponding EC model fulfills
the second principle. Exploiting the aforementioned equivalence, the
following results are derived too: (i) a rigorous thermodynamic
definition of the plastic strain and of the dissipated energy for BW
models exhibiting strength and stiffness degradation; (ii) a new
elasto-damaged BW model; (iii) a bound for the stiffness degradation
rate of BW models, proving that the traditional semi-empirical
stiffness degradation laws, like the energetic linear rule or the
so-called pivot rule, may lead to some pathological thermodynamic
behavior (these situations are illustrated by several numerical
examples); and (iv) an improved energetic stiffness degradation rule
and a new stiffness degradation rule.

After the introduction in the first Section, the existing BW models
are reviewed in the first three parts of the second Section, while a
new class of stress-strain BW models with stiffness degradation is
defined in the fourth part. The equivalence between the flow rules
of BW models with both degradations and the EC theory with damage is
discussed in the third Section. In the fourth Section, single
degree-of-freedom (1-DoF) and two-degrees-of-freedom (2-DoF) BW
structural models are analyzed in detail, using the theoretical
tools previously developed. This analysis is supplemented by some
numerical examples and by an application to partial-strength
beam-to-column steel joints.

\section{Bouc-Wen-type models}

\label{SecBW models}

A classification of Bouc-Wen-type models is suggested in Figure \ref%
{scheme_BW}. The main distinction concerns non-degrading models
(ND-BW models) and degrading ones. Moreover, among degrading models,
the further differentiation between D-BW models having only strength
degradation and models characterized by strength and stiffness
degradation or by stiffness degradation only (DD-BW models) is
pointed out. Then, the models formulated for describing the material
behavior by stress-strain laws and the structural behavior by
generalized force-displacement laws are distinguished. Note that the
box corresponding to \emph{stress-strain} DD-BW models represents a
new class of models, defined by equation (\ref{bouc type}). In this
paper, the equivalence between EC models with isotropic damage
(DD-EC), as defined in \citet{ErlicherPointA}, and the stress-strain
DD-BW models is proven and then used to provide a thermodynamically
well-posed formulation for the DD-BW models.

The stress-strain models described in this paper are presented only for a
plastically incompressible behavior, since it is the most important case in
the literature concerning stress-strain BW models. The treatment of the more
general case, with a non-zero hydrostatic plastic flow, is conceptually
similar, but requires a more complex formalism not needed here.

\subsection{ND-BW models for structures}

\label{SecNDBW modelstruct}

Among the different models of hysteresis proposed by \citet{Bouc71}, the
simplest one can be formulated by a Stieltjes integral as follows:
\begin{equation}
\left\{
\begin{array}{l}
w\left( t\right) =A_{0}\textrm{ }u\left( t\right) +Z\left( t\right) \\
Z\left( t\right) =\int_{0}^{\xi _{s}\left( t\right) }\mu \left( \xi
_{s}\left( t\right) -\xi _{s}^{\prime }\right) \textrm{ }du\left(
\xi
_{s}^{\prime }\right)%
\end{array}
\right.  \label{eqdin2}
\end{equation}
where $A_{0}$ is a non-negative constant; $u$ and $w$ are two time-dependent
functions, which are considered as input and output functions, respectively.
In structural engineering applications the input usually has the meaning of
a generalized displacement, while the output $w$ plays the role of a
generalized force, defined as the sum of a linear term $A_{0}u(t)$ and a
hysteretic term $Z(t)$. The integral in (\ref{eqdin2})$_{2}$ depends on the
time-function $\xi _{s}\left( t\right) $, which is named \emph{internal time}
and is assumed to be positive and non-decreasing. The function $\mu $,
called the \emph{hereditary kernel}, takes into account hysteretic
phenomena. One of the definitions of $\xi _{s}$ proposed by Bouc is the
total variation of $u$:
\begin{equation}
\xi _{s}\left( t\right) =\int_{0}^{t}\left\vert \frac{du}{dt^{\prime
}} \right\vert dt^{\prime }\textrm{ \ \ or, equivalently, \ \
}\dot{\xi} _{s}=\left\vert \dot{u}\right\vert \textrm{, \ with }\xi
_{s}\left( 0\right) =0 \textrm{.}  \label{dtheta_du}
\end{equation}
where the superposed dot indicates time-differentiation.
(\ref{dtheta_du}) implies the rate-independence of $\xi _{s}$ and as
a result, $Z$ and $w$ are in turn rate-independent. \citet{Bouc71}
defined $\mu $ as a continuous, bounded, positive and non-increasing
function on the interval $\xi _{s}\geq 0 $, having a bounded
integral. In the special case of an exponential kernel
\begin{equation}
\mu \left( \xi _{s}\right) =Ae^{-\beta _{s}\xi _{s}}\textrm{ \ \ \ \
\ \ \ \ \ \ \ \ \ \ \ \ \ \ with \ \ \ }A,\beta _{s}>0  \label{kern}
\end{equation}%
a differential formulation of (\ref{eqdin2}) can be easily deduced.
Then, (\ref{dtheta_du}) and (\ref{kern}) imply:
\begin{equation}
\left\{
\begin{array}{l}
w=A_{0}\textrm{ }u+Z \\
\dot{Z}=A\textrm{ }\dot{u}-\beta _{s}\textrm{ }Z\textrm{
}\dot{\xi}_{s}\textrm{ \ \ \ \ \ \ \ \ with
}\dot{\xi}_{s}=\left\vert \dot{u}\right\vert
\end{array}
\right.  \label{eqbouc}
\end{equation}
One can observe that for an initial value in the interval $\left(
-Z_{u},Z_{u}\right) $, with $Z_{u}=A/\beta _{s}$, the hysteretic
force $Z$ remains in the same interval. \citet{Bouc67} proposed a
more general formulation of (\ref{eqbouc}):
\begin{equation}
\left\{
\begin{array}{l}
w=A_{0}\textrm{ }u+Z \\
\dot{Z}=A\textrm{ }\dot{u}-\beta _{s}\textrm{ }Z\textrm{ }\left\vert
\dot{u} \right\vert -\gamma _{s}\left\vert Z\right\vert
\dot{u}\textrm{ \ \ \ \ \ \ \
\ \ \ \ \ \ with \ \ }\gamma _{s}<\beta _{s}%
\end{array}
\right.  \label{Boucbetgam}
\end{equation}
while \citet{Wen76} suggested a further modification by introducing a
positive exponent $n$:
\begin{equation}
\left\{
\begin{array}{l}
w=A_{0}\textrm{ }u+Z \\
\dot{Z}=A\textrm{ }\dot{u}-\left( \beta _{s}\textrm{ }sgn\left(
Z\textrm{ }\dot{u} \right) +\gamma _{s}\right) \left\vert
Z\right\vert ^{n}\dot{u}
\end{array}
\right.  \label{wen}
\end{equation}
where $sgn(\cdot )$ is the signum function. Wen did not impose any
condition on the value of $\gamma _{s}$ and assumed integer values
for $n$; however, all real positive values of $n$ are admissible.
When $n$ is large enough, force-displacement curves similar to those
of an elastic-perfectly-plastic model with the additional linear
term $A_{0}u$ are obtained. Provided that $\beta _{s}+\gamma
_{s}>0,$ the limit strength value $Z_{u}$ of (\ref{wen}) becomes:
\begin{equation}
Z_{u}=\left( \frac{A}{\beta _{s}+\gamma _{s}}\right) ^{\frac{1}{n}}
\end{equation}%
The parameter $\beta _{s}$ is positive by assumption, while the
admissible values for $\gamma _{s}$ can be derived from the
condition $\dot{\xi}_s \geq 0$ \citep{ErlicherPoint2004}. See also
Table \ref{tabRiass1}, in this respect.

\subsection{Plastically incompressible ND-BW stress-strain models}

\label{SecNDBW models}

In order to link the classical plasticity theory and the Bouc-Wen model
described in (\ref{wen}), \citet{Karray89} proposed a tensorial
generalization of (\ref{wen}) for isotropic and plastically incompressible
materials:
\begin{equation}
\left\{
\begin{array}{l}
tr\left( \mbox{\boldmath $\sigma$}\right) =3E_{1}\textrm{ }tr\left(
\mbox{\boldmath $\varepsilon$}\right) +3K\textrm{ }tr\left(
\mbox{\boldmath
$\varepsilon$}\right) \\
dev\left( \mbox{\boldmath $\sigma$}\right) =2E_{2}\textrm{
}dev\left(
\mbox{\boldmath $\varepsilon$}\right) +\mathbf{z} \\
\mathbf{\dot{z}}=2G\textrm{ }dev\left( \mbox{\boldmath
$\dot{\varepsilon}$}\right) -\beta \textrm{ }\mathbf{z}\textrm{
}\dot{\xi}
\end{array}
\right.  \label{Casciati}
\end{equation}
with $\beta >0$ and the following special choice for the intrinsic time
\begin{equation}
\dot{\xi}=\left( 1+\frac{\gamma }{\beta }\textrm{ }sgn\left(
\mathbf{z:}dev\left( \mbox{\boldmath $\dot{\varepsilon}$}\right)
\right) \right) \left\vert \mathbf{z}:dev\left( \mbox{\boldmath
$\dot{\varepsilon}$}\right) \right\vert \left\Vert
\mathbf{z}\right\Vert ^{n-2}  \label{csiKBC}
\end{equation}%
The terms proportional to $E_{1}\geq 0$ and $E_{2}\geq 0$ play the same role
as $A_{0}$. $\mbox{\boldmath
$\varepsilon$}$ and $\mbox{\boldmath
$\sigma$}$ are the second-order symmetric strain and stress tensors,
respectively; $tr$ and $dev$ are the trace and the deviatoric operator,
respectively; $\mathbf{z}$ is a traceless tensor defining the hysteretic
part of the stress, while $\left\Vert \mathbf{\cdot }\right\Vert $ is the
standard $L_{2}-norm$ of a symmetric second-order tensor. $K$ and $G$ are
the bulk modulus and the shear modulus, respectively. Note that the
stress-strain relationship for the traces is linear, according to the
assumption of plastic incompressibility. \citet{Casciati89} proposed a
stochastic analysis of the same model. For this reason, we name it as
Karray-Bouc-Casciati (KBC) model. The norm of the tensor $\mathbf{z}\left(
t\right) $ is bounded as follows:
\begin{equation}
\left\Vert \mathbf{z}\right\Vert =\left\Vert dev\left(
\mbox{\boldmath $\sigma$}\right) -2E_{2}\textrm{ }dev\left(
\mbox{\boldmath $\varepsilon$} \right) \right\Vert < \sigma
_{u}=\left( \frac{2G}{\beta +\gamma }\right) ^{\frac{1}{n}}
\end{equation}
for $t>0$, provided that $\left\Vert \mathbf{z}\left( 0\right)
\right\Vert <\sigma _{u}$. This inequality shows that a limit
strength value exists for the KBC model and only regards the
deviatoric part of the stress $\mbox{\boldmath $\sigma$}$,
consistently with the plastic incompressibility requirement. A
thermodynamic formulation of this model, based on a suited
definition of the Helmholtz free energy and of a pseudo-potential,
was proposed in \citet{ErlicherPoint2005}.

\subsection{DD-BW models for structures}

\label{SecDDBW modelstruct}

\begin{figure}[tbp]
\begin{center}
\includegraphics[width=14cm]{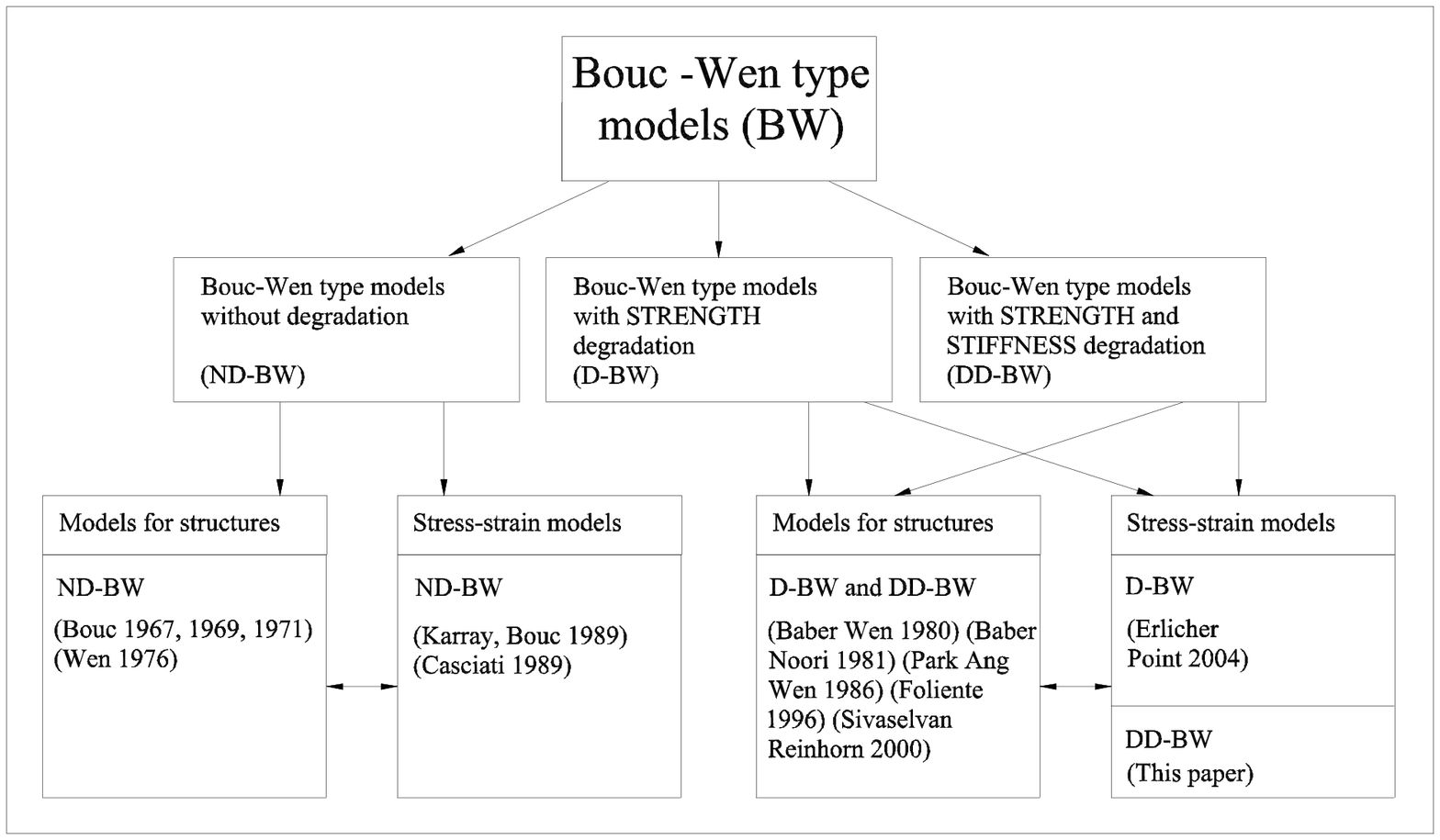}
\end{center}
\caption{Classification of Bouc-Wen-type models.}
\label{scheme_BW}
\end{figure}

The BW model (\ref{wen}) was modified by \citet{Baber81}, who introduced the
positive and increasing functions $\eta _{s}$ and $\nu _{s}$, both having a
unit initial value:
\begin{equation}
\left\{
\begin{array}{l}
w=A_{0}\textrm{ }u+Z \\
\dot{Z}=\frac{1}{\eta _{s}}\left[ A\textrm{ }\dot{u}-\nu _{s}\left(
\beta _{s} \textrm{ }sgn\left( Z\textrm{ }\dot{u}\right) +\gamma
_{s}\right) \left\vert Z\right\vert ^{n}\textrm{ }\dot{u}\right]
\end{array}
\right.  \label{BaberWen}
\end{equation}
where $\nu _{s}$ represents a \emph{strength degradation} effect,
while $\eta _{s}$ is associated with a \emph{stiffness degradation}
effect. According to Figure \ref{scheme_BW}, an attentive reader can
see that this model belongs to the class of DD-BW models for
structures. A degradation function similar to $\eta _{s}$ could also
be associated with the \emph{post-yielding} stiffness $A_{0}$.
However, in order to simplify the present analysis, $A_0$ is
supposed to be constant. In the original formulation of
\citet{Baber81}, $\eta _{s}$ and $\nu _{s}$ are defined as linearly
increasing functions of $e_{s}$, the energy dissipated by the
hysteretic model:
\begin{equation}
\nu _{s}=1+c_{\nu ,s}\textrm{ }e_{s}\ ,\ \ \ \ \ \eta _{s}=1+c_{\eta
,s}\textrm{ }e_{s}  \label{rules}
\end{equation}%
with $c_{\nu ,s}\geq 0$ and $c_{\eta ,s}\geq 0$. The authors did not provide
any definition of $e_{s}$ on a thermodynamic basis. This topic will be
discussed later, in the fourth Section. The case of \emph{increasing}
strength, viz. $0<\nu _{s}\leq 1$, is also admissible and it is analogous to
isotropic hardening in classical plasticity models. Provided that $\beta
_{s}+\gamma _{s}>0$, the maximum strength value, modified by the strength
coefficient $r$, becomes
\begin{equation}
Z_{u,red}=\left( \frac{A}{\nu _{s}\textrm{ }\left( \beta _{s}+\gamma
_{s}\right) }\right) ^{\frac{1}{n}}=\left( \frac{1}{\nu _{s}}\right)
^{\frac{1}{n}}Z_{u}=r\ Z_{u}
\end{equation}
Several applications of this version of the BW model can be found in
the literature; see, among others, \citet{Foliente95} and
\citet{Sivaselvan2000}. The latters proposed a formulation of
(\ref{BaberWen}) entailing a clear physical meaning of the
parameters. Moreover, the following degradation rules were
suggested:
\begin{equation}
\left\{
\begin{array}{l}
\nu _{s}=\nu _{s}\left( e_{s},u_{\max }\right) =\left( 1-\left( \frac{%
u_{\max }}{u_{ult}}\right) ^{1/\beta _{1}}\right) ^{-n}\left( 1-\frac{\beta
_{2}}{1-\beta _{2}}\frac{e_{s}}{e_{ult}}\right) ^{-n} \\
\eta _{s}=\eta _{s}\left( u,w\right) =\frac{A}{R_{k}\left(
A+A_{0}\right) -A_{0}}=\frac{A}{\frac{w+\alpha \textrm{
}\tilde{Z}_{u,red}}{\left( A+A_{0}\right) u+\alpha \textrm{
}\tilde{Z}_{u,red}}\left( A+A_{0}\right)
-A_{0}}%
\end{array}%
\right.  \label{rulesSiva}
\end{equation}%
where $u_{\max }=\max_{0\leq \tau \leq t}\left( \left\vert u\left( \tau
\right) \right\vert \right) $ is the actual value of the maximum
displacement modulus; $u_{ult}$ and $e_{ult}$ are defined as the ultimate
displacement and the ultimate dissipated energy, corresponding to failure; $%
\beta _{1}$ and $\beta _{2}$ are positive parameters related to strength
degradation; $\alpha $ is a positive parameter related to the stiffness
degradation rule (\ref{rulesSiva})$_{2}$, called the \emph{pivot} rule. The
sign of $\tilde{Z}_{u,red}=Z_{u,red}$ $sgn\left( u-\frac{w}{\left(
A+A_{0}\right) }\right) $ is related to the position of the actual
force-displacement point $\left( u,w\right) $ with respect to the line
representing the initial elastic behavior $w=\left( A+A_{0}\right) u$.

In summary, the general univariate (also called uniaxial or 1-DoF)
BW model reads:
\begin{equation}
\left\{
\begin{array}{l}
w=A_{0}\textrm{ }u+Z \\
\dot{Z}=\frac{1}{\eta _{s}}\left( A\textrm{ }\dot{u}-\beta
_{s}\textrm{ }Z\textrm{
}\nu _{s}\dot{\xi}_{s}\right)%
\end{array}%
\right.  \label{1D}
\end{equation}%
where $\eta _{s},\nu _{s}$ and $\dot{\xi}_{s}$ can vary from one model
to another, as shown in Table \ref{tabRiass1}.
%The expressions of
%$\dot{\xi}_{s}$ for BW models with strength degradation only (D-BW
%models) were first given by \citet{ErlicherPoint2004}.

\begin{figure}[tbp]
\begin{center}
\includegraphics[width=14cm]{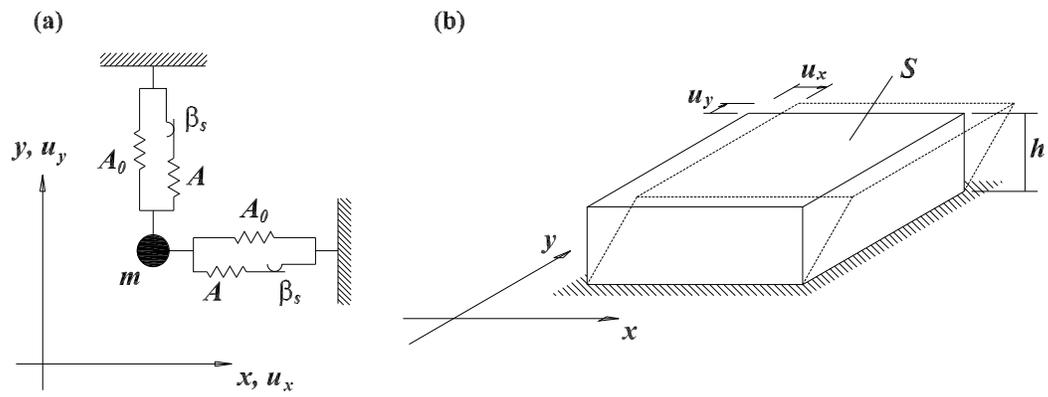}
\end{center}
\caption{(a) Scheme of the system associated with the 2-DoF model
(\protect \ref{2D}). \emph{m} is the system mass. (b) An ideal
bearing device subjected to bi-directional shear excitation.}
\label{Fig_ShearDevice}
\end{figure}

A 2-DoF generalization of (\ref{1D}) was defined by \textbf{\
}\cite{Park et al 1986} to represent the behavior of a system
constituted of a single mass \emph{m} subjected to an excitation
acting in two orthogonal directions. A simple model associated with
this situation is illustrated in Figure \ref{Fig_ShearDevice}a. For
instance, this model is suited to reproduce the geometrically linear
uncoupled behavior of a bi-axially loaded reinforced concrete column
\citep{Park et al 1986} or the uncoupled biaxial behavior of a
laminated rubber bearing \citep{Abe et al 2004}. In the simplest
case where the stiffness and the strength are the same in both
directions, i.e. isotropic behavior, one has
\begin{equation}
\left\{
\begin{array}{l}
\mathbf{W}=A_{0}\mathbf{U+Z} \\
\mathbf{\dot{Z}=}\frac{1}{\eta _{s}}\left( A\mathbf{\dot{U}}-\beta
_{s}\mathbf{Z}\textrm{ }\nu _{s}\dot{\xi}_{s}\right)
\end{array}
\right.   \label{2D}
\end{equation}
where $\mathbf{W=}\left( w_{x},w_{y}\right) $ is the total force
vector, $\mathbf{U=}\left( u_{x},u_{y}\right) $ is the structural
displacement and $\mathbf{Z=}\left( Z_{x},Z_{y}\right) $ is the
hysteretic force; the parameters $A_{0}$ and $A$ have the dimension
of a stiffness; $\eta _{s}$\textbf{,} $\nu _{s}$ are scalars, since
the damage and strength evolution are assumed to be equal in both
directions. The intrinsic time
$\dot{\xi}_{s}=\dot{\xi}_{s}\left(Z_{x},\dot{u}_{x},Z_{y},\dot{u}_{y}\right)$
and the degradation functions $\nu_s$, $\eta_s$ may be chosen
according to one of the definitions given in Table
\ref{tabRiass1_1}.

An equivalent way of writing (\ref{2D}) is
\begin{equation}
\left\{
\begin{array}{l}
w_{x}=A_{0}\textrm{ }u_{x}+Z_{x} \\
w_{y}=A_{0}\textrm{ }u_{y}+Z_{y} \\
\dot{Z}_{x}=\frac{1}{\eta _{s}}\left( A\textrm{ }\dot{u}_{x}-\beta
_{s}\nu
_{s}Z_{x}\dot{\xi}_{s}\right)  \\
\dot{Z}_{y}=\frac{1}{\eta _{s}}\left( A\textrm{ }\dot{u}_{y}-\beta
_{s}\nu _{s}Z_{y}\dot{\xi}_{s}\right)
\end{array}%
\right.   \label{2D_expl}
\end{equation}%
A special case of (\ref{2D_expl}) is given by \citep{Park et al
1986}
\begin{equation}
\left\{
\begin{array}{l}
w_{x}=A_{0}\textrm{ }u_{x}+Z_{x} \\
w_{y}=A_{0}\textrm{ }u_{y}+Z_{y} \\
\dot{Z}_{x}=\frac{1}{\eta _{s}}\left( A\textrm{ }\dot{u}_{x}-\nu
_{s}\left[ \beta _{s}\textrm{ }Z_{x}\left\vert
Z_{x}\dot{u}_{x}\right\vert +\gamma _{s} \textrm{
}Z_{x}^{2}\dot{u}_{x}+\beta _{s}\textrm{ }Z_{x}\left\vert
Z_{y}\dot{u}
_{y}\right\vert +\gamma _{s}\textrm{ }Z_{x}Z_{y}\dot{u}_{y}\right] \right)  \\
\dot{Z}_{y}=\frac{1}{\eta _{s}}\left( A\textrm{ }\dot{u}_{y}-\nu
_{s}\left[ \beta _{s}\textrm{ }Z_{y}\left\vert
Z_{y}\dot{u}_{y}\right\vert +\gamma _{s} \textrm{
}Z_{y}^{2}\dot{u}_{y}+\beta _{s}\textrm{ }Z_{y}\left\vert
Z_{x}\dot{u} _{x}\right\vert +\gamma _{s}\textrm{
}Z_{x}Z_{y}\dot{u}_{x}\right] \right)
\end{array}
\right.   \label{ParkAngWenDispl}
\end{equation}
which corresponds to a particular choice of the intrinsic time
$\dot{\xi}_{s} $, as indicated in the second row of Table
\ref{tabRiass1_1}. \citet{Park et al 1986} assumed $\beta
_{s}=\gamma _{s}$ with $A_{0}$ and $\gamma _{s}$ having a special
coupled evolution, introduced to obtain force-displacement curves
confined in a given envelope. However, we shall assume that $A_{0}$
and $\gamma _{s}$ are constant. The strength degradation $\nu _{s}$
was defined as in (\ref{rules})$_{1}$, while the stiffness
degradation $\eta _{s}$ was given by:
\begin{equation}
\eta _{s}=\eta _{s}\left( u_{\max ,x},u_{\max ,y}\right) =\sqrt{\eta
_{x}^{2}+\eta _{y}^{2}}=\sqrt{\left( \frac{u_{\max ,x}}{u_{yield}}\right)
^{2}+\left( \frac{u_{\max ,y}}{u_{yield}}\right) ^{2}}\geq 1
\label{rulesPAW}
\end{equation}%
where $u_{yield}$ is the yielding displacement, $u_{\max ,x}$ and
$u_{\max ,y}$ are the actual maximum values of the displacement in
the two directions $x$ and $y$. The stiffness degradation is
therefore governed by the "uniaxial" ductility ratios $\eta _{x}$
and $\eta _{y}$, while $\eta _{s}$ is a "bi-axial" ductility ratio.

\subsection{Plastically incompressible DD-BW stress-strain models}

\label{SecDDBW models}

The KBC model (\ref{Casciati})-(\ref{csiKBC}) is a stress-strain BW model
characterized by the absence of strength and stiffness degradation. In this
subsection, a direct generalization of (\ref{Casciati}) is proposed, leading
a DD-BW model:
\begin{equation}
\left\{
\begin{array}{l}
tr\left( \mbox{\boldmath $\sigma$}\right) =3E_{1}\textrm{ }tr\left(
\mbox{\boldmath $\varepsilon$}\right) +\frac{1}{\eta }3K\textrm{
}tr\left(
\mbox{\boldmath $\varepsilon$}\right)  \\
dev\left( \mbox{\boldmath $\sigma$}\right) =2E_{2}\textrm{
}dev\left(
\mbox{\boldmath $\varepsilon$}\right) +\mathbf{z} \\
\mathbf{\dot{z}}=\frac{1}{\eta }\left( 2G\textrm{ }dev\left(
\mbox{\boldmath $\dot{\varepsilon}$}\right) -\beta \textrm{
}\mathbf{z}\textrm{ }\nu \textrm{ } \dot{\xi}\right)
\end{array}
\right.   \label{bouc type}
\end{equation}
The linearity of the hydrostatic stress-strain law is due to the
assumption of plastic incompressibility.
The thermodynamic
admissibility of the DD-BW model defined in (\ref{bouc type}), i.e.
the conditions under which it fulfills the second principle of
thermodynamics, is discussed in the two following Sections. Some
simple degradation functions may be defined by analogy with the
structural energetic linear rules discussed in the previous
subsection, viz.
\begin{equation}
\nu =1+c_{\nu }\textrm{ }e\ ,\ \ \ \ \ \eta =1+c_{\eta }\textrm{ }e
\label{rulesTens}
\end{equation}%
where $e$ is the dissipated \ energy per unit volume and the
coefficients $c_{\nu },$ $c_{\eta }$ are analogous to $c_{\nu ,s},$
$c_{\eta ,s}$ introduced in (\ref{rules}). The expressions of $\eta
,\ \nu $ and $\dot{\xi}$ for different models are collected in Table
\ref{tabRiass2}. As already stated for structural models, the case
of \emph{increasing} strength, viz. $0<\nu \leq 1$, is also
admissible and it is analogous to isotropic hardening in classical
plasticity.

Models with strength degradation lead to stress-strain laws with
softening and, if applied in the continuum setting, to strain
localization. It is well-known that an objective description of
softening materials and the resulting localized failure modes
require special attention. Thus, some of the model parameters must
be considered not as pure material properties but as
discretization-dependent or, alternatively, the underlying theory
must be enriched by special terms, like weighted spatial averages or
higher-order gradients, acting as localization limiters \citep[Chap.
21]{JirasekBazant02}. These aspects are not the main subject of this
paper and therefore are not treated here.

\section{Equivalence between DD-BW stress-strain models and endochronic
models with damage}

The endochronic models with isotropic damage (DD-EC) are discussed
in a companion paper \citep{ErlicherPointA}. We are interested here
in the analysis of a DD-EC stress-strain law having the same format
as the force-displacement rules (\ref{1D}) and (\ref{2D}), which is
defined by the following relationships
\begin{equation}
\left\{
\begin{array}{l}
\mbox{\boldmath $\sigma$}=\mathbf{E}:\mbox{\boldmath
$\varepsilon$}+\left( 1-D\right) \mathbf{C:}\left(
\mbox{\boldmath
$\varepsilon -\varepsilon$}^{p}\right) \\
tr\left( \mbox{\boldmath $\dot{\varepsilon}$}^{p}\right) =0\textrm{
\ \ \ \ \ \ \ and \ \ \ \ \ \ \ \ \ }\mbox{\boldmath
$\dot{\varepsilon}$}^{p}=\frac{1}{1-D}\frac{dev\left( \mathbf{\sigma
-E:\varepsilon }\right) }{2G/\beta }\ \frac{\dot{\zeta}}{g}%
\end{array}%
\right.  \label{DD-EC}
\end{equation}%
The non-degrading linear term
$\mathbf{E}:\mbox{\boldmath$\varepsilon$}$ is analogous to the term
$A_0 u$ characterizing the models (\ref{1D}) and (\ref{2D}). The
tensor $\mbox{\boldmath$\varepsilon$}^{p}$ is the plastic strain.
The trace of the plastic strain flow is zero because of the plastic
incompressibility assumption; $\mathbf{C}=\left(
K-\frac{2G}{3}\right) \mathbf{1\otimes 1}+2G\mathbf{I}$ is the
standard elasticity fourth-order tensor for isotropic materials;
$\mathbf{E}=\left( E_{1}-\frac{2E_{2}}{3} \right) \mathbf{1\otimes
1}+2E_{2}\mathbf{I}$ is a fourth-order tensor defining an additional
elastic stiffness; $\mathbf{1}$ is the second-order identity tensor;
$\mathbf{I}$ is the fourth order identity tensor and
$\mathbf{\otimes }$ represents the tensor product; $D$ is the scalar
damage variable, supposed to be non-negative and less than one;
$\zeta $ is called the \emph{intrinsic time measure} and
$\dot{\zeta}\geq 0$ has the role of plastic multiplier
\citep{ErlicherPointA}; $g>0$ is sometimes called the
\emph{hardening-softening function} \citep{Bazant78}. (\ref {DD-EC})
is equivalent to
\begin{equation}
\left\{
\begin{array}{l}
tr\left( \mathbf{\mbox{\boldmath $\sigma$}}\right) =3E_{1}\textrm{
}tr\left( \mbox{\boldmath $\varepsilon$}\right) +\left( 1-D\right)
3K\textrm{ }tr\left( \mbox{\boldmath
$\varepsilon$}\right) \\
dev\left( \mbox{\boldmath $\sigma$}\right) =2E_{2}\textrm{
}dev\left( \mbox{\boldmath
$\varepsilon$}\right) +\mathbf{z} \\
\mathbf{\dot{z}}=\left( 1-D\right) \textrm{ }2G\textrm{ }dev\left(
\mbox{\boldmath $\dot{\varepsilon}$}\right) -\beta \textrm{
}\mathbf{z}\textrm{ }
\frac{\dot{\zeta}}{g}-\dot{D}\textrm{ }\frac{\mathbf{z}}{1-D}%
\end{array}%
\right.  \label{sigDam}
\end{equation}%
where the time derivative of the hysteretic deviatoric tensor
$\mathbf{z}$ is explicitly written\textbf{.} The comparison of
(\ref{bouc type}) and (\ref {sigDam}), in particular the definition
of $\mathbf{\dot{z}}$, shows that the two models, DD-BW and DD-EC,
are equivalent if
\begin{equation}
\nu =\frac{1}{g}\ ,\ \ \ \ \eta =\frac{1}{1-D}\ ,\ \ \ \
\dot{\xi}=\frac{\dot{\zeta}}{1-D}+\dot{D}\textrm{ }\frac{g}{\beta
\left( 1-D\right) ^{2}} \label{EquivBW-Endo}
\end{equation}
with $\nu \left( 0\right) =1,$ $D\left( 0\right) =0$ and $\xi \left(
0\right) =\zeta \left( 0\right) =0$. (\ref{EquivBW-Endo}) relate the
functions $D$, $g$ and $\zeta $ associated with the DD-EC model and
the functions $\eta $, $\nu $ and $\xi $ associated with the DD-BW
model. Obviously, this equivalence assumes that the evolution of
$D$, $g$ and $\zeta $ can be defined in a way that is general enough
to represent all possible evolution of the given $\eta $, $\nu $ and
$\xi $. To be more precise, if for instance $\eta $ and $\nu $ are
proportional to the dissipated energy, which for DD-BW models may
increase during unloading phases, then it should be possible to
define $g$ and $D$ of the equivalent endochronic model such that
they depend on the dissipated energy, with possible non-zero
increments also during unloading phases. The proper thermodynamic
framework for DD-EC models allowing this general behavior is
presented in a companion paper \citep{ErlicherPointA}.
(\ref{EquivBW-Endo})$_{1}$ shows that the function $\nu $ defining
the strength increment (degradation) for BW models is strictly
related to the function $g$ defining the isotropic hardening
(softening) for endochronic models. (\ref{EquivBW-Endo})$_{2}$ shows
a similar equivalence between the stiffness degradation function of
DD-BW models and the isotropic damage variable for DD-EC models. The
interpretation of (\ref{EquivBW-Endo})$_{3}$ is more involved. It
states that the intrinsic time flow $\dot{\xi}$ of a DD-BW model can
be expressed as the sum of: (i) a contribution related to the
plastic multiplier $\dot{\zeta}$ of the equivalent DD-EC model; (ii)
a part related to damage flow $\dot{D}$ of the equivalent DD-EC
model. Let us set (\ref{EquivBW-Endo})$_{3}$ in the alternative form
\begin{equation}
\dot{\zeta}=\frac{\dot{\xi}}{\eta }-\frac{\dot{\eta}}{\beta \eta \nu }
\label{equivBW-endo1}
\end{equation}%
One can see that (\ref{equivBW-endo1}) is the \emph{definition of
the plastic multiplier} $\dot{\zeta}$\emph{\ of the DD-EC model
equivalent to a given DD-BW model, as a function of} $\beta
,$\emph{\ }$\dot{\xi},$\emph{\ }$\eta $\emph{\ and }$\nu $. Briefly,
(\ref{equivBW-endo1}) defines the \emph{plastic multiplier of
stress-strain DD-BW models}. When $\dot{\zeta}=0$, there is no
plastic flow, viz. an \emph{elastic with damage} behavior is
retrieved. Moreover, $\dot{\zeta}=0$ if and only if
\begin{equation}
\dot{\eta}=\frac{\dot{D}}{\left( 1-D\right) ^{2}}=\beta \nu \dot{\xi}
\label{stiffSpecial}
\end{equation}%
i.e. for DD-BW models, it is possible to obtain zero plastic
strains, provided that the special stiffness degradation rule
(\ref{stiffSpecial}) is adopted. Figure \ref{Fig_ElDam11} shows this
behavior, with the parameters $E1=E2=0,$ $2G=24000$ $MPa$, $K=20000$
$MPa$, $\dot{\xi}$ is defined by (\ref{csiKBC}) with $n=1.2$, $\beta
=81.2409$ $MPa^{1-n}$, $\gamma /\beta =0.5$ (entailing
$\sigma_{u}=100\sqrt{(2/3)}$ $MPa$), $\nu =1+c_{\nu }e$ with $c_{\nu
}=0.05$ $m^{3}MJ^{-1}$, $\dot{e}$ is given in (\ref{disEnerg})
underafter and $\dot{\eta}$ is defined by (\ref{stiffSpecial}).

\section{Thermodynamic analysis of DD-BW models}

\label{SecBWfinal}

\subsection{DD-BW stress-strain models}

The thermodynamic analysis of the endochronic model with isotropic damage
(DD-EC) presented by \citet{ErlicherPointA} led to the following conditions
ensuring thermodynamic admissibility:
\begin{equation}
\begin{array}{ll}
\dot{D}\geq 0 & \textrm{: the damage is non-decreasing} \\
\dot{\zeta}\geq 0 & \textrm{: the plastic multiplier is
non-negative}
\end{array}
\label{thermoCondEndo}
\end{equation}
These conditions derive from the assumption that the dissipated
energy increments owing to plasticity and to damage are
\emph{separately} non-negative. It is well known that this
assumption is not needed, since only the \emph{total} dissipated
energy must be non-decreasing. However, this assumption suffices to
define models fulfilling the second principle; it is usually adopted
\citep{Lemaitre90engl} and is also used hereafter.
(\ref{EquivBW-Endo})$_{2}$ and (\ref{thermoCondEndo})$_{1}$ lead to
\begin{equation}
\dot{\eta}\geq 0\textrm{ : the stiffness degradation function is
non-decreasing}  \label{damCond1}
\end{equation}
while (\ref{EquivBW-Endo})$_{3}$ and (\ref{thermoCondEndo})$_{1-2}$ entail
\begin{equation}
\dot{\xi}\geq 0\textrm{ : the intrinsic time of BW models is
non-decreasing} \label{csiCond}
\end{equation}
For DD-BW models, an expression of $\dot{\xi}$ fulfilling
(\ref{csiCond}) is usually given, while $\dot{\zeta}$ is unknown. It
follows that the positivity of $\dot{\zeta}$, recall
(\ref{thermoCondEndo})$_2$, must be ensured by a relevant condition
on the stiffness degradation rule, which can easily be derived using
(\ref{equivBW-endo1}):
\begin{equation}
\begin{array}{l}
\dot{\eta}\leq \beta \textrm{ }\nu \ \dot{\xi}\textrm{ : upper bound
for the
stiffness degradation rate}%
\end{array}
\label{damCond}
\end{equation}%
In summary, two conditions on the stiffness degradation rule are
provided: the first one, i.e. (\ref{damCond1}), is trivial, while
the second one, (\ref{damCond}), is less obvious. They are obtained
using (\ref{EquivBW-Endo}) and (\ref{equivBW-endo1}), that relate a
DD-BW model, defined by $\dot{\xi}$, $\eta $ and $\nu $, with the
\emph{equivalent} DD-EC model, defined by $\dot{\zeta}$, $D$ and
$g$.

\begin{figure}[tbp]
\begin{center}
\includegraphics[width=14cm]{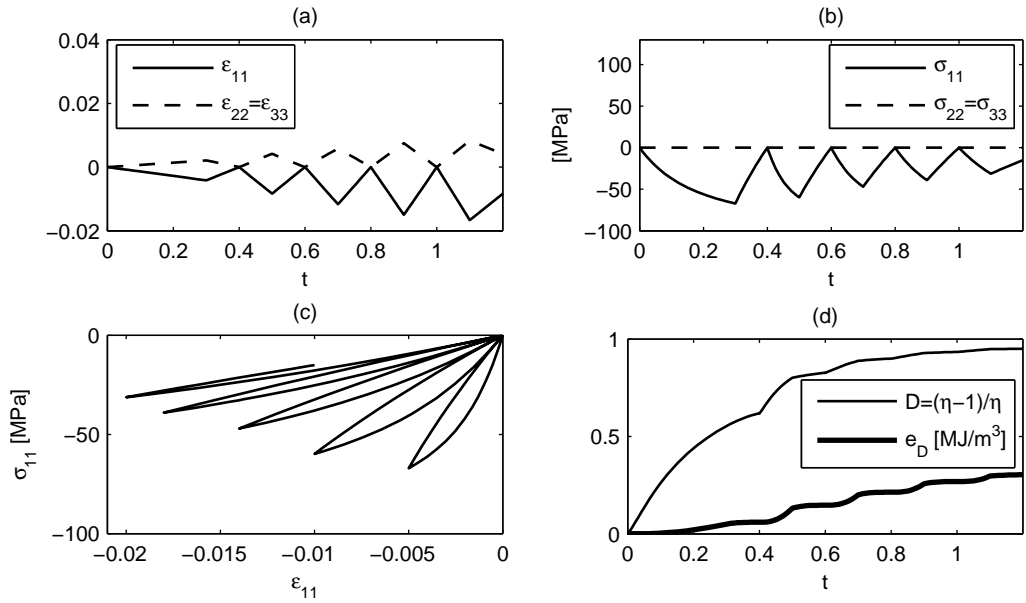}
\end{center}
\caption{Elasto-damaged BW model, uniaxial stress $\protect\sigma_{11}=\frac{
3}{2}[dev(\protect\sigma)]_{11}$. (a) Strain-history $\protect\varepsilon%
_{11}$ and $\protect\varepsilon _{22}$. (b) $\protect\sigma_{11}$ stress
evolution. (c) Stress-strain loops. (d) The damage evolution with the
corresponding dissipated energy.}
\label{Fig_ElDam11}
\end{figure}

The inequality (\ref{damCond}) suggests the definition of a new stiffness
degradation rule (recall that $\eta \geq 1$):
\begin{equation}
\dot{\eta}=c_{\beta }\left( \frac{1}{\eta }\right) ^{m}\textrm{
}\beta \textrm{ }\nu \textrm{ }\dot{\xi}  \label{stiffNew}
\end{equation}
with $c_{\beta }\in \left[ 0,1\right] $ and $m\geq 0$. For $m=0$,
(\ref{stiffNew}) entails (see also
(\ref{EquivBW-Endo})-(\ref{equivBW-endo1}))
\begin{equation}
\dot{\eta}=c_{\beta }\textrm{ }\beta \textrm{ }\nu \textrm{
}\dot{\xi}\textrm{ \ \ } \Longleftrightarrow \textrm{\ \ \ \
}\dot{\zeta}=\left( 1-D\right) \textrm{ }\dot{\xi}\left( 1-c_{\beta
}\right)
\end{equation}
i.e. the plastic multiplier is proportional to $\left( 1-D\right) $: with
the increase of damage, the plastic multiplier decreases. In other words,
this rule postulates that plastic strains are larger when the material is
slightly damaged and viceversa. When also $c_{\beta }=1$ holds, then $\dot{%
\zeta}=0$ and an \emph{elastic with damage} behavior is retrieved, as shown
in Figure \ref{Fig_ElDam11}. %\ref{Fig_ElDam} and

\begin{figure}[tbp]
\begin{center}
\includegraphics[width=14cm]{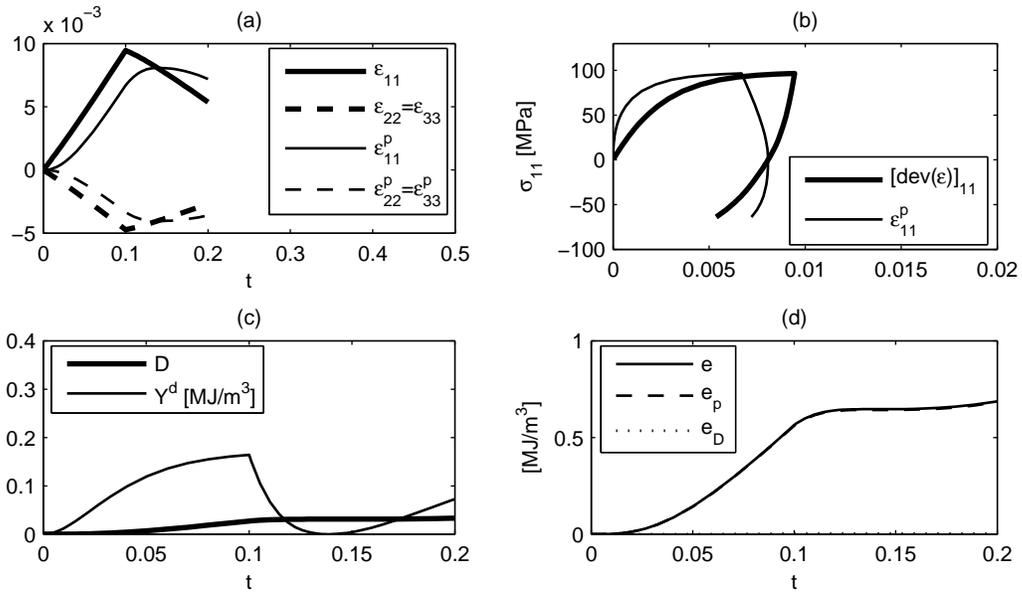}
%% 14cm * 8cm, carattere minimo 8pt
\end{center}
\caption{Loading-unloading loop of a BW model, uniaxial stress $\protect%
\sigma_{11}=\frac{3}{2}[dev(\protect\sigma )]_{11}$. (a) Total and plastic
strains. (b) Stress-strain behavior. (c) The evolution of $Y^{d}$ and of the
damage. (d) Dissipated energies.}
\label{Fig_defPlast3}
\end{figure}

Moreover, one can observe that, still using (\ref{EquivBW-Endo})-(\ref%
{equivBW-endo1}) and recalling that $\mathbf{z=}dev\left(
\mbox{\boldmath
$\sigma$}-\mathbf{E}:\mbox{\boldmath$\varepsilon$}\right) $, the energies
dissipated by plasticity and by damage read:
\begin{equation}
\begin{array}{l}
\dot{e}=\dot{e}_{p}+\dot{e}_{D}\textrm{ \ \ \ \ \ \ \ with} \\
\dot{e}_{p}=\left( \mathbf{\mbox{\boldmath $\sigma$ }}-\mathbf{E}:
\mbox{\boldmath$\varepsilon$}\right) :\mbox{\boldmath
$\dot{\varepsilon}$}^{p}=\beta
\frac{\mathbf{z}:\mathbf{C}^{-1}:\mathbf{z}}{ \left( 1-D\right)
}\left( \left( 1-D\right) \nu \textrm{ }\dot{\xi}-\frac{\dot{
D}}{\beta \left( 1-D\right) }\right)
=\frac{\mathbf{z}:\mathbf{z}}{2G}\left(
\beta \nu \dot{\xi}-\dot{\eta}\right)  \\
\dot{e}_{D}=Y^{d}\dot{D}=\frac{1}{2}\left( \mathbf{\mbox{\boldmath
$\sigma$ } }-\mathbf{E}:\mbox{\boldmath$\varepsilon$}\right)
:\mathbf{C}^{-1}:\left( \mathbf{\mbox{\boldmath $\sigma$
}}-\mathbf{E}:\mbox{\boldmath$\varepsilon$} \right)
\frac{\dot{D}}{\left( 1-D\right) ^{2}}=\frac{1}{2}\left( \frac{
\mathbf{z}:\mathbf{z}}{2G}+\frac{\left( tr\left( \sigma
-E\varepsilon \right) \right) ^{2}}{9K}\right) \dot{\eta}
\end{array}
\label{disEnerg}
\end{equation}
See also the corresponding definitions for DD-EC models in
\citet{ErlicherPointA}. $Y^d$ is the dissipative part of the
thermodynamic force dual of the damage variable $D$. Note that the
increments of the energy dissipated by plasticity $\dot{e}_{p}$
depend not only on the BW intrinsic time flow $\dot{\xi}$ but also
on the flow $\dot{\eta}$ of the stiffness degradation function.
Figure \ref{Fig_defPlast3} shows the evolution of the two terms of
the dissipated energy as a function of time. The parameter values
are the same of those of Figure \ref{Fig_ElDam11}, except for the
ratio $\gamma /\beta =-0.5$ and $\beta =243.7227$ $MPa^{1-n}$; $\nu
=1+c_{\nu }e$ with $c_{\nu }=0.05$ $ m^{3}MJ^{-1}$; $\eta =1+c_{\eta
}e_{p}$ with $c_{\eta }=0.05$ $m^{3}MJ^{-1}$
. %The energy increments $\dot{e}$ and
%$\dot{e_{p}}$ are given in (\ref{disEnerg}).
Note that there are non-zero plastic strain increments during
unloading phases. Moreover, damage increments are always non-zero,
also during unloading phases; see also \citet{ErlicherPointA} in
this respect.
%For comparison, the case $\gamma=\beta$ is shown
%in Figure \ref{Fig_defPlast4}. The value $\beta=60.9307$ $MPa^{1-n}$
%is chosen to have the same value of $\sigma_u$ as before. The other
%parameters are the same of Figure \ref{Fig_defPlast3}.

\subsection{DD-BW models for structures}

\begin{figure}[tbp]
\begin{center}
\includegraphics[width=14cm]{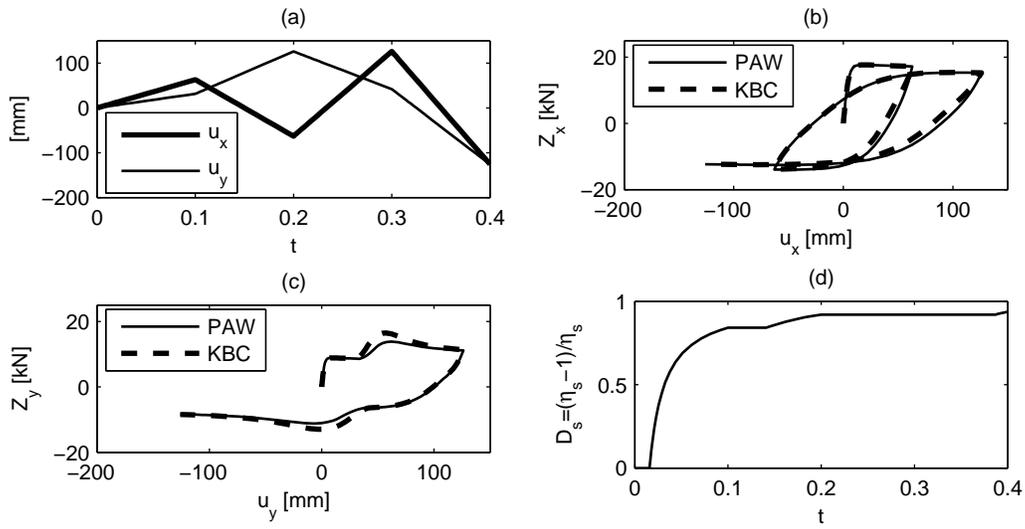}
%%% 14 x 7.0 cm
\end{center}
\caption{\emph{Non-proportional} loading of 2-DoF PAW and KBC
models. (a) Displacement history for the x- and y-directions. (b)
Force-displacement loops $(u_x,Z_x)$. (c) Force-displacement loops
$(u_y,Z_y)$. (d) Damage evolution: the curve is the same for both
models; see (\ref{rulesPAW}).} \label{FigPAWvsCascNonPROP}
\end{figure}

Comparing (\ref{bouc type}) (that is equivalent to (\ref{DD-EC})
under the conditions previously discussed) with (\ref{1D}) and
(\ref{2D}), one can see that structural models have the same format
as stress-strain models, provided that stress is replaced by force
and strain by displacement, respectively. Analogous substitutions
are made for the stiffness tensor, the $\beta ,\gamma $ parameters,
the degradation functions and the intrinsic time (see the first two
columns of Table \ref{Table-Micro-Macro}). In the ideal case of a
bearing device subjected to bi-directional shear, see \citet{Abe et
al 2004} and Figure \ref{Fig_ShearDevice}b, the relationships
between the stress-strain and force-displacement parameters are
given in the third column of Table \ref{Table-Micro-Macro}. In
general, if a direct derivation of structural models from
stress-strain models is not possible, the formal analogy between
stress-strain and force-displacement rules must be
\emph{postulated}. Accounting for this analogy and referring to the
isotropic 2-DoF model (\ref{2D}), we assume a force-displacement
relationship similar to the stress-strain law (\ref{DD-EC}), that
reads:
\begin{equation}
\mathbf{W:}=A_{0}\mathbf{U+Z}=A_{0}\mathbf{U+}\frac{1}{\eta _{s}}A\left(
\mathbf{U-U}^{p}\right) .  \label{Wa0U}
\end{equation}
The seventh line of Table \ref{Table-Micro-Macro}, with the
definition of $D_{s}$, has also been used. Moreover, from
(\ref{2D}), one has
\begin{equation}
\mathbf{\dot{W}}=A_{0}\mathbf{\dot{U}}+\frac{1}{\eta _{s}}\left[
A\mathbf{\dot{U}}-\beta _{s}\mathbf{Z}\textrm{ }\nu _{s}\textrm{
}\dot{\xi}_{s}\right] \label{2D_1}
\end{equation}
Equation (\ref{2D_1}) and the time-derivative of (\ref{Wa0U}) lead to the
definition of the \emph{plastic displacement flow}
\begin{equation*}
\mathbf{\dot{U}}^{p}=\frac{\mathbf{Z}}{A}\left( \beta _{s}\nu
_{s}\dot{\xi} _{s}-\dot{\eta}_{s}\right) =\frac{\mathbf{Z}}{A}\beta
_{s}\nu _{s}\eta _{s}\left( \frac{\dot{\xi}_{s}}{\eta
_{s}}-\frac{\dot{\eta}_{s}}{\beta _{s}\nu _{s}\eta _{s}}\right)
:=\frac{\mathbf{Z}}{A}\beta _{s}\nu _{s}\eta _{s}\dot{\zeta}_{s}
\end{equation*}%
where $\dot{\zeta}_{s}\geq 0$ is a \emph{structural plastic
multiplier}; compare with (\ref{equivBW-endo1}). By analogy with
(\ref{disEnerg}) and using Table \ref{Table-Micro-Macro}, the rate
of energy dissipated by the force-displacement model can be defined
as
follows:%
\begin{equation}
\begin{array}{l}
\dot{e}_{s}=\dot{e}_{s,p}+\dot{e}_{s,D}\textrm{ \ \ \ \ \ \ \ with} \\
\dot{e}_{s,p}=\mathbf{Z}^{T}\mathbf{\cdot
\dot{U}}^{p}=\frac{\mathbf{Z}^{T} \mathbf{\cdot Z}}{A}\textrm{
}\left( \beta _{s}\nu _{s}\dot{\xi}_{s}-\dot{\eta}
_{s}\right)  \\
\dot{e}_{s,D}:=\frac{\mathbf{Z}^{T}\mathbf{\cdot
Z}}{2A}\frac{\dot{D}_{s}}{ \left( 1-D_{s}\right)
^{2}}=\frac{\mathbf{Z}^{T}\mathbf{\cdot Z}}{2A}\textrm{ }
\dot{\eta}_{s}
\end{array}
\label{dissEnergiesBW}
\end{equation}%
All the previous quantities are defined for a 2-DoF BW model; 1-DoF
models correspond to deal with scalar quantities instead of vectors.
Concerning 2-DoF models, it is interesting to analyze the
relationship between the case of an intrinsic time of the
Park-Ang-Wen (PAW) type (see the second row of Table
\ref{tabRiass1_1}), versus a 2-DoF model having an intrinsic time of
the KBC type \emph{with $n=2$} (see the last row of Table
\ref{tabRiass1_1}). It is easy to prove that the intrinsic time
flows $\dot{\xi}_{s}$ of the two models become identical in the case
of \emph{proportional loading}. However, they \emph{are different in
the case of non-proportional loading}, leading to different
hysteretic loops, as illustrated in Figure
\ref{FigPAWvsCascNonPROP}. The parameters chosen for the numerical
simulation are $n=2$, $A_{0}=0$, $A=3.571$ $kN/mm$, $\beta
_{s}=0.0119$ $kN^{1-n}mm^{-1}$, $\gamma _{s}=-\beta _{s}/4$.
Moreover, the strength degradation is defined by (\ref{rules})$_{1}$
with $c_{\nu,s}=0.0001$ $J^{-1}$, while the stiffness degradation by
(\ref{rulesPAW}) with $u_{yield}=11.2$ $mm$.

\subsubsection{Discussion about the stiffness degradation rule}

Conditions (\ref{damCond1}) and (\ref{damCond}) hold at the stress-strain
law level. However, taking into account the above-mentioned analysis, the
same constraints are \emph{postulated} at the force-displacement level. As a
result, one has
\begin{equation}
\dot{\eta}_{s}\geq 0\ ,\ \ \ \ \dot{\eta}_{s}\leq \beta _{s}\textrm{
}\nu _{s} \textrm{ }\dot{\xi}_{s}  \label{damCond1s}
\end{equation}
Likewise, also the stiffness degradation rule (\ref{stiffNew}) can be
extended to BW models for structures:
\begin{equation}
\dot{\eta}_{s}=c_{\beta }\left( \frac{1}{\eta _{s}} \right)
^{m}\textrm{ }\beta _{s}\textrm{ }\nu _{s}\textrm{ }\dot{\xi}_{s}\ \
\ \ \textrm{with}\ \ c_{\beta }\in \left[ 0,1\right] \ \textrm{and}\
\ m\geq 0 \label{stiffNew1}
\end{equation}

According to the proposed analysis, every stiffness degradation rule
defined for structural BW models should be checked with respect to
(\ref{damCond1s}). As a result, some bounds for the parameters
characterizing these degradation rules may be found. Two examples
are considered here: the energetic linear rule, where a possible
violation of (\ref{damCond1s})$_{2}$ is highlighted, and the pivot
rule, where (\ref{damCond1s})$_{1}$ may not be fulfilled.

%\subsubsection{Example 1: the energetic linear rule}

\begin{figure}[tbp]
\begin{center}
\includegraphics[width=17.78cm]{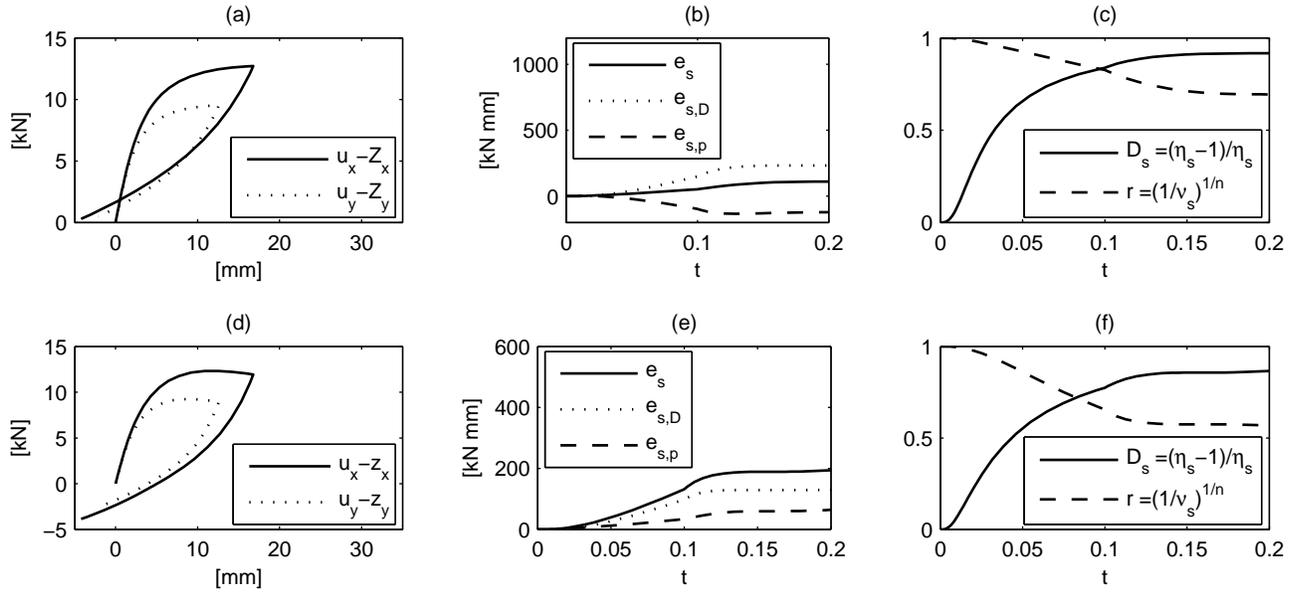}
%%% 17.78 x 8.0 cm
\end{center}
\caption{2-DoF Bouc-Wen model with energetic linear degradation
rules. Case of stiffness degradation proportional to $e_{s}$: (a)
loading-unloading loop; (b) dissipated energies. Note that $e_{s,p}$
is negative; (c) damage evolution and strength reduction. Case of
stiffness degradation proportional to $e_{s,p}$: (d)
loading-unloading loop; (e) dissipated energies; (f) damage
evolution and strength reduction.} \label{Fig_DegrEn1}
\end{figure}

The \emph{energetic linear rule} defined in (\ref{rules})$_{2}$,
first adopted in \citet{Baber81}, postulates that the stiffness
degradation function $\eta _{s}$ depends linearly on the dissipated
energy $e_{s}=e_{s,p}+e_{s,D}$, defined in (\ref{dissEnergiesBW}):
\begin{equation}
\dot{\eta}_{s}=c_{\eta ,s}\ \dot{e}_{s}=c_{\eta ,s}\left(
\dot{e}_{s,p}+\dot{ e}_{s,D}\right) \ \ \rightarrow
\dot{\eta}_{s}=\frac{c_{\eta ,s}\frac{ \mathbf{Z}^{T}\mathbf{\cdot
Z}}{A}}{1+c_{\eta ,s}\frac{1}{2}\frac{\mathbf{Z} ^{T}\mathbf{\cdot
Z}}{A}}\beta _{s}\nu _{s}\dot{\xi}_{s}  \label{ineqStiff0}
\end{equation}
As a result, (\ref{damCond1s})$_{2}$ imposes that the coefficient
multiplying $\beta _{s}\nu _{s}\dot{\xi}_{s}$ in
(\ref{ineqStiff0})$_{2}$ be less or equal than one, entailing
$c_{\eta ,s}\mathbf{\ Z}^{T}\mathbf{\cdot Z}\leq 2A$. This is an
upper bound for the parameter $c_{\eta ,s}\geq 0$, varying with the
actual value of $\mathbf{Z}$. When it is violated,
(\ref{damCond1s})$_{2}$ is not satisfied, leading to a negative
plastic multiplier, i.e. $\dot{\zeta}_{s}<0$, and to negative
increments of the energy dissipated by plasticity, i.e.
$\dot{e}_{s,p}<0$. This situation is considered as non-admissible,
since it is assumed that the energies dissipated by plasticity and
damage must be \emph{separately} increasing. A numerical simulation
where this pathological situation occurs is presented in Figures
\ref{Fig_DegrEn1}a-c, where a 2-DoF BW model with intrinsic time of
KBC type is considered; in detail, $A_{0}=0$, $A=3.5714$ $kN/mm$,
$\beta _{s}=0.1308$ $kN^{1-n}mm^{-1}$, $\gamma _{s}=-\beta _{s}/4$
and $n=1.2$. The strength degradation is governed by
(\ref{rules})$_{1}$ with $c_{\nu ,s}=0.005$ $J^{-1}$, while the
stiffness degradation is given by (\ref{rules})$_{2}$ with $c_{\eta
,s}=0.1$ $J^{-1}$. Conversely, defining a stiffness degradation rule
that depends linearly on the energy dissipated \emph{by plasticity}
\begin{equation}
\begin{array}{l}
\eta _{s}=1+c_{\eta ,s}e_{s,p} \\
\dot{\eta}_{s}=c_{\eta ,s}\textrm{ }\dot{e}_{s,p}=c_{\eta
,s}\frac{\mathbf{Z} ^{T}\mathbf{\cdot Z}}{A}\left( \beta _{s}\nu
_{s}\dot{\xi}_{s}-\dot{\eta} _{s}\right) \textrm{\ \ \ }\rightarrow
\textrm{ \ \ }\dot{\eta}_{s}=\frac{ c_{\eta
,s}\frac{\mathbf{Z}^{T}\mathbf{\cdot Z}}{A}}{1+c_{\eta ,s}\frac{
\mathbf{Z}^{T}\mathbf{\cdot Z}}{A}}\beta _{s}\nu _{s}\dot{\xi}_{s}
\end{array}
\label{StiffEnPlast}
\end{equation}
the limit condition (\ref{damCond1s})$_{2}$ is always fulfilled.
Hence, for a stiffness degradation proportional to the
\emph{plastic} dissipated energy, no limitations are needed on
$c_{\eta ,s}$. The numerical example illustrated in Figures
\ref{Fig_DegrEn1}d-f confirms this result. All the model parameters
are the same used for the previous case.

%\subsubsection{Example 2: the "pivot" rule}

\begin{figure}[tbp]
\begin{center}
\includegraphics[width=17.78cm]{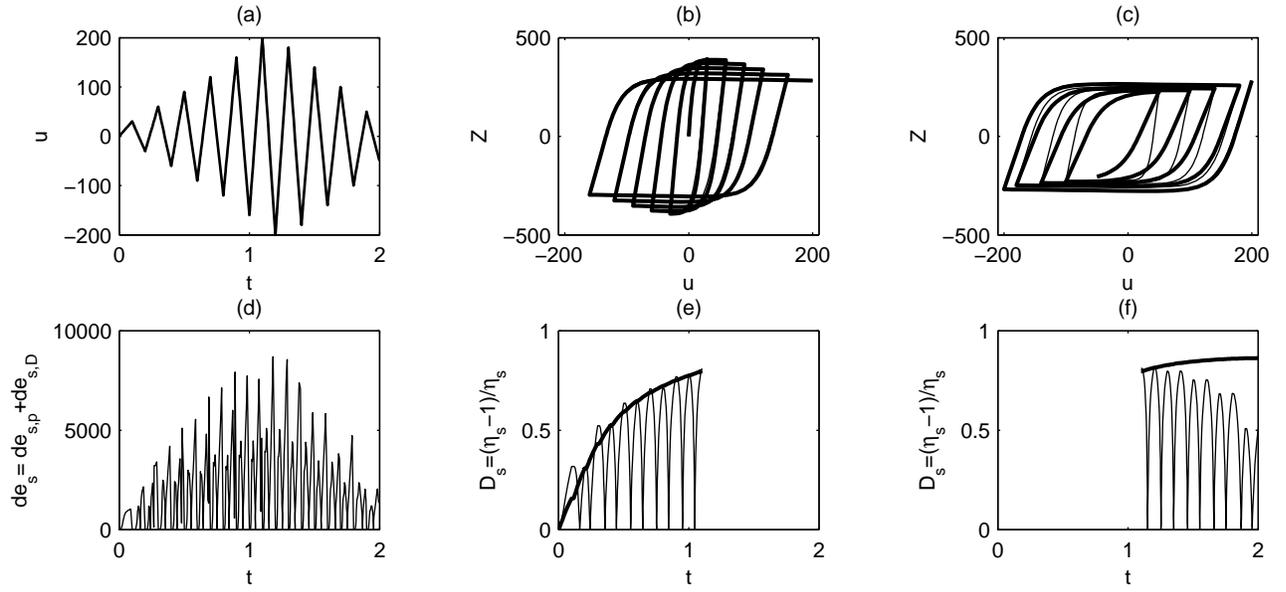}
%%% 17.78 x 8.0 cm
\end{center}
\caption{Pivot rule (thin line) vs. the new stiffness degradation
rule (\protect\ref{stiffNew1})(thick line). (a) Displacement
history. (b) Force-displacement loops for the increasing amplitude
phase and (c) for the decreasing amplitude phase. (d) Dissipated
energy increments for the pivot rule. (e) Damage evolutions for the
increasing amplitude phase and (f) for the decreasing amplitude
phase.} \label{FigPivot_vs_New}
\end{figure}

The \emph{pivot rule} (\ref{rulesSiva})$_{2}$ is sometimes used for
modeling the cyclic behavior of structures; see e.g.
\citet{Sivaselvan2000} in this respect. Some problems associated
with this rule, occurring when it is applied with displacements
histories with cycles of \emph{decreasing} amplitude, were pointed
out by \citet{WangFoliente2001}. A BW model is considered here, with
$A_{0}=0$, $A=50$, $\gamma _{s}=\beta _{s}=0.0189$, $n=1.2$
(entailing $Z_{u}=400$) and $\nu _{s}=1+c_{\nu ,s}e_{s}$ with
$c_{\nu ,s}=1e-6$. The parameter $\alpha $ of the pivot rule is
equal to $5$. In this example, the same difficulties mentioned by
\citet{WangFoliente2001} appear, viz. the stiffness \emph{increases}
when the cycle amplitude decreases: see Figure
\ref{FigPivot_vs_New}c. Moreover, a deeper problem is also
highlighted: even if the energy dissipated \emph{by plasticity} and
the \emph{total} dissipated energy are non-decreasing, and the
second principle is fulfilled, see Figure \ref{FigPivot_vs_New}d,
this rule may entail \emph{negative} damage increments, \emph{both}
during increasing and decreasing amplitude loops. As a result,
$\dot{D}_{s}<0\Longleftrightarrow \dot{\eta}_{s}<0$ with a violation
of (\ref{damCond1s})$_{1}$, as shown in Figures
\ref{FigPivot_vs_New}e,f.

\begin{figure}[tbp]
\begin{center}
\includegraphics[width=14cm]{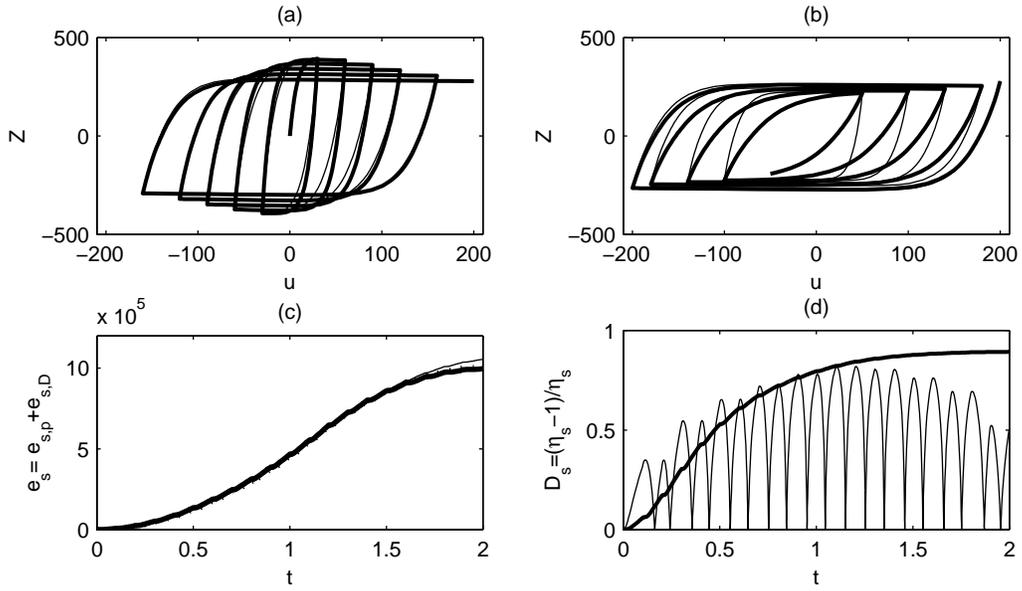}
%%% 14 x 8.0 cm
\end{center}
\caption{Pivot rule (thin line) vs. the stiffness degradation rule
(\protect\ref{StiffEnPlast}), based on the dissipated energy by
plasticity (thick line). (a) Force-displacement loops for the
increasing amplitude phase and (b) for the decreasing amplitude
phase. (c) Total dissipated energy by the pivot rule and by the
plastic energy rule. The dotted line indicates the total energy
dissipated in the case of the stiffness rule
(\protect\ref{stiffNew1}), shown in Figure
\protect\ref{FigPivot_vs_New}. (d) Damage evolution.}
\label{FigPivot_vs_enp}
\end{figure}

Let us now compare the pivot rule to the stiffness degradation rule
(\ref{stiffNew1}). Figure \ref{FigPivot_vs_New}b shows that it is
possible to choose the parameters for both rules in such a way to
have almost identical force-displacement loops during an initial
phase characterized by increasing displacement amplitude cycles. In
the specific case considered in the Figure, the new stiffness
degradation rule (\ref{stiffNew1}) is characterized by $c_{\beta
}=0.06$ and $m=1.3$. The damage evolution during this phase for both
degradation laws is depicted in Figure \ref{FigPivot_vs_New}e: the
new degradation rule entails a monotonic increase of damage,
consistent with the assumption that dissipated energies by
plasticity and damage \emph{separately} increase; while the pivot
rule is characterized by phases with alternate increase and decrease
of damage. Then, decreasing amplitude loops are considered for both
models and they provide very different behaviors, as illustrated in
Figure \ref{FigPivot_vs_New}f: the rule (\ref{stiffNew1}) still
entails an increase of damage, while the pivot rule induces an
averaged decrease of damage, corresponding to an averaged increase
of stiffness. A similar comparison can be made between the pivot
rule and (\ref{StiffEnPlast}). Figure \ref{FigPivot_vs_enp} shows
the numerical results with $c_{\eta ,s}=8.4e-6.$ The same remarks
made for Figure \ref{FigPivot_vs_New} hold: Figure
\ref{FigPivot_vs_enp}d shows the evolution of the total dissipated
energy for the case of the pivot rule (thin line), the rule
(\ref{StiffEnPlast}) (thick line) and the rule (\ref{stiffNew1})
considered in the previous example (dotted line). As expected, the
pivot rule has a different evolution in the last part of the
displacement history, where the stiffness increases instead of
decreasing like for the other two rules.

\begin{figure}[tbp]
\begin{center}
\includegraphics[width=5cm]{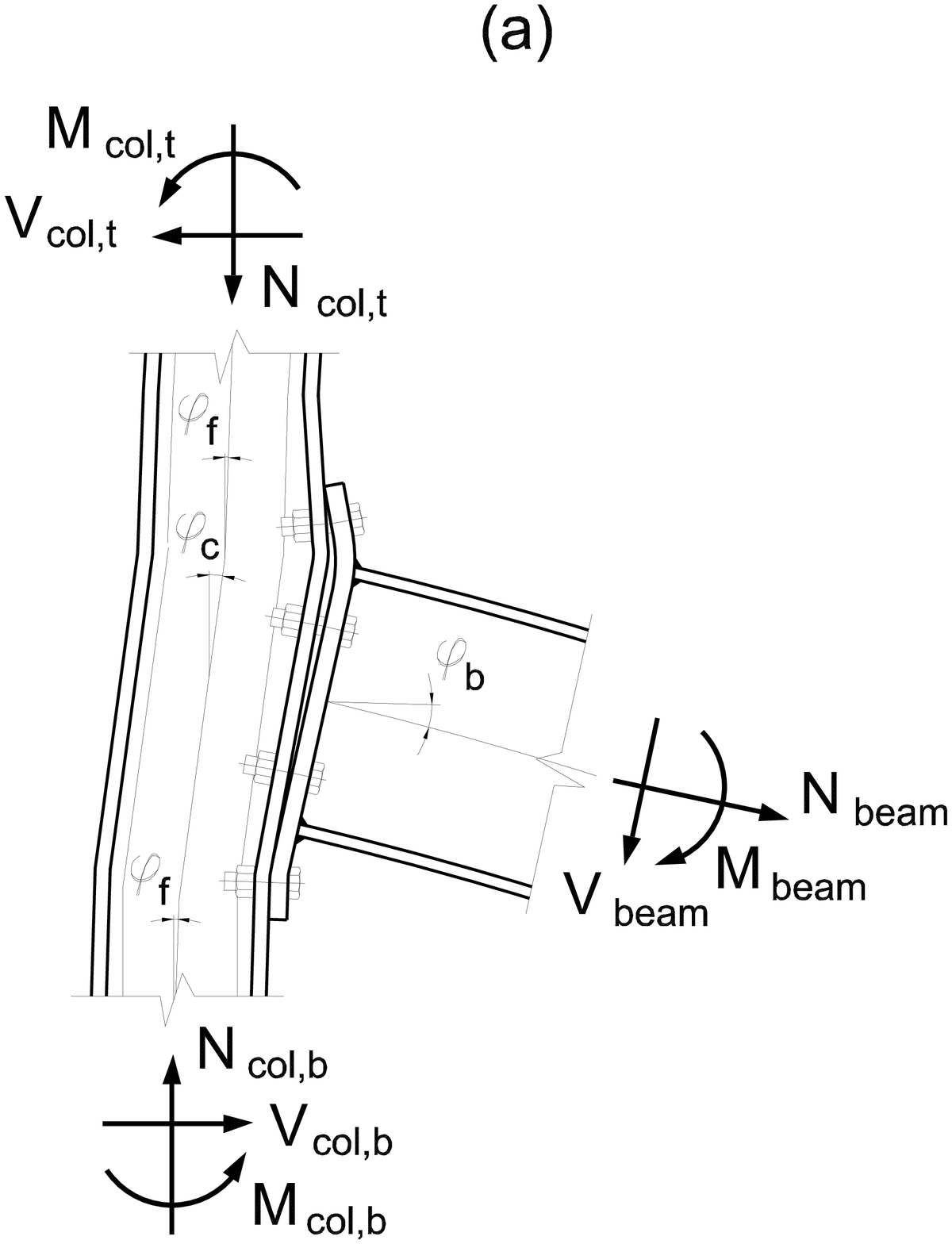} %%% 5 x 7.0 cm
\includegraphics[width=11cm]{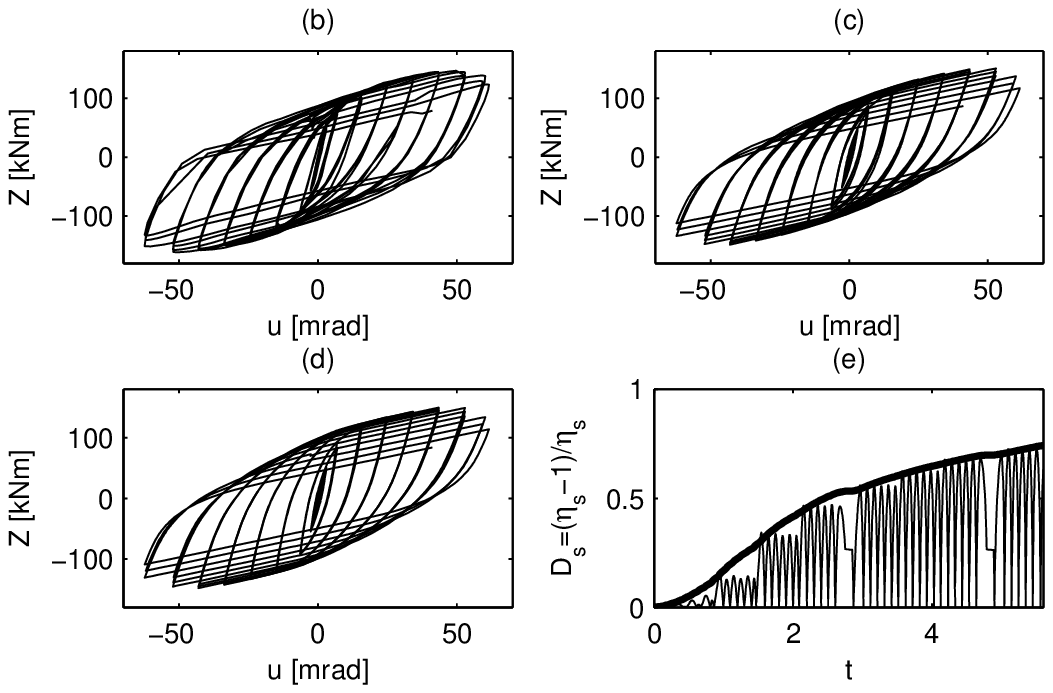} %%% 11 x 7.0 cm
\end{center}
\caption{Application to experimental data. (a) Partial strength
beam-to-column steel joint JB1-3A \citep{bursietal2002}. (b)
Experimental Moment-rotation loops: u=$\varphi_b-\varphi_f$,
Z=M$_{beam}$ (c) Predicted loops with the rule (\ref{stiffNew1}).
(d) Predicted loops with the pivot rule (\ref{rulesSiva})$_2$. (e)
Damage evolution predicted by the model endowed with the rule
(\ref{stiffNew1}) (thick line) and by the model with the pivot rule
(\ref{rulesSiva})$_2$ (thin line).} \label{Fig_Exp}
\end{figure}

An application application of the BW model (\ref{BaberWen}) with the
stiffness degradation (\ref{stiffNew1}) and the strength degradation
$\nu _{s}=\left( 1-\left( \frac{e_{s}}{e_{ult}}\right)
^{\frac{1}{\beta _{3}}}\right) ^{-n}$, inspired to the rule
(\ref{rulesSiva})$_{1}$ is shown in Figure \ref{Fig_Exp}. In detail,
the experimental loops refer to a partial-strength beam-to-column
steel joint, studied by \citet{bursietal2002}. The assumed parameter
values are $A_{0}=1.0$ $kNm/mrad$, $A=21.0$ $kNm/mrad$, $\beta
_{s}=0.4888$ $\left( kNm\right)^{1-n}\left( mrad\right) ^{-1}$,
$\gamma _{s}=0,$ $n=0.8$, $Z_{u}=110$ $kNm$, $e_{ult}=210000$ $Nm$,
$\beta _{3}=0.35$, $c_{\beta }=0.02$ and $m=1.8$. For the pivot
rule, we assumed $\alpha=4$. An attentive reader can observe that a
good agreement between experimental (Figure \ref{Fig_Exp}b) and
numerical (Figure \ref{Fig_Exp}c,d) hysteretic loops is obtained. In
addition, the evolution of damage predicted by the model with the
rule (\ref{stiffNew1}) and with the pivot rule is illustrated in
Figure \ref{Fig_Exp}e. Again, the drawback of the pivot rule is
evident. In summary, one can state that (\ref{stiffNew1}) and
(\ref{StiffEnPlast}) may be used as a valid alternative to the pivot
rule to predict accurate loops; furthermore, they represent a
significant improvement in terms of thermodynamic admissibility.

\section{Conclusions}

An exhaustive classification of Bouc-Wen-type models was provided in
this paper. The models endowed with both strength and stiffness
degradation were compared with an extended endochronic theory with
isotropic damage. The thermodynamic formulation of this extended
endochronic theory was discussed in a companion paper
\citep{ErlicherPointA}, using pseudo-potentials depending on state
variables and on parameters related to the past history of the
material. Using this theoretical framework, a rigorous thermodynamic
definition of the energy dissipated by plasticity and damage was
provided for Bouc-Wen models having both strength and stiffness
degradation, as well as the definition of an elastic with damage
Bouc-Wen model. Moreover, an important constraint on the stiffness
degradation rules for Bouc-Wen models was highlighted and its
consequences on the energetic linear rule were investigated. The
non-monotonic damage evolution characterizing the pivot rule was
illustrated. Finally, an improved formulation of the energetic
linear rule and a new stiffness degradation rule were proposed;
their effectiveness with respect to the pivot rule was proved by
numerical examples and an application to beam-to-column steel
joints.

\begin{ack}
The authors would like to thank prof. Nelly Point and acknowledge
the several discussions which helped the improvement of this work.
The authors are also grateful to a reviewer for his constructive
criticism.
\end{ack}

\section{Appendix: Notations}

\textit{The following symbols are used in this paper:}

$A_{0}=$ additional stiffness for structural BW models

$A=$ initial stiffness for structural BW models

$\mathbf{C}=$ fourth-order elasticity tensor

$D=$ internal variable associated with isotropic damage

$\mathbf{E}=$ additional fourth-order elasticity tensor

$E_{1}=$ bulk modulus of the additional elastic behavior

$E_{2}=$ shear modulus of the additional elastic behavior

$K=$ bulk modulus

$G=$ shear modulus

$\mathbf{I}=$ fourth-order identity tensor

$Y^{d}=$ dissipative part of the thermodynamic force dual of the damage
variable

$\mathbf{1}=$ second-order identity tensor

\bigskip

$e=$ total dissipated energy per unit volume

$e_{s}=$ total structural dissipated energy

$e_{p}=$ energy per unit volume dissipated by plasticity

$e_{s,p}=$ structural energy dissipated by plasticity

$e_{D}=$ energy per unit volume dissipated by damage

$e_{s,D}=$ structural energy dissipated by damage

$g=$ hardening-softening function for EC models

$u=$ total displacement for 1-DoF BW models

$\mathbf{U}=$ total displacement vector for 2-DoF BW models

$w=$ total force for 1-DoF BW models

$\mathbf{W}=$ total force vector for 2-DoF BW models

$Z=$ hysteretic force for 1-DoF BW models

$\mathbf{z}=$ hysteretic part of the stress tensor

$\mathbf{Z}=$ hysteretic force vector for 2-D0F BW models

$\alpha =$ coefficient associated with the pivot degradation rule

$\beta =$ coefficient characterizing the flow rules of BW or EC
stress-strain models

$\beta _{s}=$ coefficient characterizing the flow rules of structural BW
models

$\gamma =$ additional coefficient defining the intrinsic time of BW or EC
stress-strain models

$\gamma _{s}=$ additional coefficient defining the intrinsic time of
structural BW models

$\mbox{\boldmath $\varepsilon$}=$ total small strain tensor

$\mbox{\boldmath $\varepsilon$}^{p}=$ plastic strain tensor

$\mu =$ hereditary kernel

$\nu =$ strength degradation function for BW stress-strain models

$\nu_s =$ strength degradation function for BW structural models

$\eta =$ stiffness degradation function for BW stress-strain models

$\eta_s =$ stiffness degradation function for BW structural models

$\mbox{\boldmath $\sigma$}=$ Cauchy stress tensor

$\vartheta =$ intrinsic time \emph{scale} for EC models

$\xi =$ intrinsic time for BW stress-strain models

$\xi _{s}=$ intrinsic time for 1-DoF and 2-DoF BW models

$\zeta =$ intrinsic time \emph{measure} for EC models; plastic multiplier
for stress-strain BW models

$\zeta _{s}=$ structural plastic multiplier for 1-DoF and 2-DoF BW models

\clearpage

%%%%%%%%%%%%%%%%%%%%%%%%%%%%%%%%%%%%%%%%%%%%%%%%%%%%%%%%%%%%%%%%%%%%%%%%%%%%%

\listoftables \clearpage

\begin{table}[tbp]
\begin{tabular}{llll}
\hline
& \vline Eq. & \vline $%
\begin{array}{l}
\nu_s ,\ \eta_s%
\end{array}
$ & \vline $\dot{\xi}_s$ \\ \hline
\citep{Bouc67} \citep{Bouc71} & \vline (\ref{1D}) & \vline $1$ , \ $1$ & %
\vline $\left\vert \dot{u }\right\vert $ \\
\citep{Bouc67} \citep{Bouc71} & \vline (\ref{1D}) & \vline $1$ \ , \ $1$ & %
\vline $\left( 1+\frac{ \gamma _{s}}{\beta _{s}}\textrm{ }sgn\left(
Z\textrm{
\textbf{\ }}\dot{u}\right) \right) \left\vert \dot{u}\right\vert $ \\
\citep{Wen76} & \vline (\ref{1D}) & \vline $1$ \ , \ $1$ & \vline
$\left( 1+ \frac{\gamma _{s}}{ \beta _{s}}\textrm{ }sgn\left(
Z\textrm{\textbf{\ }}\dot{u} \right) \right) \left\vert
\dot{u}\right\vert \left\vert Z\right\vert ^{n-1}$
, \ $n>0$ \\
$%
\begin{array}{l}
\textrm{\citep{Baber81} and} \vspace{-12pt} \\
\multicolumn{1}{c}{\textrm{\citep{Sivaselvan2000}}}%
\end{array}
$ & \vline (\ref{1D}) & \vline $%
\begin{array}{l}
\textrm{Eq. } (\ref{rules}) \vspace{-12pt} \\
\multicolumn{1}{c}{\textrm{Eq. }(\ref{rulesSiva})}%
\end{array}
$ & \vline $\left( 1+\frac{\gamma _{s}}{\beta _{s}}\textrm{ }sgn\left( Z\textrm{%
\textbf{\ }}\dot{u}\right) \right) \left\vert \dot{u}\right\vert \left\vert
Z\right\vert ^{n-1}$, \ $n>0$ \\
This paper & \vline (\ref{1D}) & \vline $%
\begin{array}{l}
\textrm{Eq. } (\ref{rules})_1, (\ref{stiffNew1}) \vspace{-12pt} \\
\multicolumn{1}{c}{\textrm{Eq. }(\ref{rulesSiva})_1, (\ref{stiffNew1})} \\
\multicolumn{1}{c}{\textrm{Eq. }(\ref{rules})_1, (\ref{StiffEnPlast})}%
\end{array}
$ & \vline $\left( 1+\frac{\gamma _{s}}{\beta _{s}}\textrm{ }sgn\left( Z\textrm{%
\textbf{\ }}\dot{u}\right) \right) \left\vert \dot{u}\right\vert \left\vert
Z\right\vert ^{n-1}$, \ $n>0$ \\ \hline
\end{tabular}%
\caption{Summary of the 1-DoF BW models for structures. $\dot{\protect\xi}%
_{s}\geq 0$, provided that $\left\vert \protect\gamma_{s}\right\vert \leq
\protect\beta_{s}$. For a thermodynamically admissible model, the stiffness
degradation rule fulfills the condition $\dot{\protect\eta}_s\leq \protect%
\beta_{s}\protect\nu_s \dot{\protect\xi}_{s}$ (see equation
(\ref{damCond1s})$_2$).} \label{tabRiass1}
\end{table}

\clearpage

\begin{table}[tbp]
\begin{tabular}{llll}
\hline
& \vline Eq. & \vline $%
\begin{array}{l}
\nu_s ,\ \eta_s%
\end{array}
$ & \vline $\dot{\xi}_s$ \\ \hline
\citep{Park et al 1986} & \vline (\ref{2D}) & \vline Eq. (\ref{rules})$_1$, (%
\ref{rulesPAW}) & \vline $\left\vert Z_{x} \textrm{
}\dot{u}_{x}\right\vert +\left\vert Z_{y}\textrm{ }\dot{u}
_{y}\right\vert +\frac{\gamma _{s}}{\beta
_{s}}\left( Z_{x}\textrm{ }\dot{u} _{x}+Z_{y}\textrm{ }\dot{u}_{y}\right) $ \\
This paper & \vline (\ref{2D}) & \vline$%
\begin{array}{l}
\textrm{Eq. } (\ref{rules}) \vspace{-12pt} \\
\textrm{Eq. }(\ref{rules})_1, (\ref{stiffNew1})%
\end{array}
$ & \vline $\left\vert Z_{x} \textrm{ }\dot{u}_{x}\right\vert
+\left\vert Z_{y} \textrm{ }\dot{u} _{y}\right\vert +\frac{\gamma
_{s}}{\beta _{s}}\left(
Z_{x} \textrm{ }\dot{u} _{x}+Z_{y}\textrm{ }\dot{u}_{y}\right) $ \\
This paper & \vline (\ref{2D}) & \vline $%
\begin{array}{l}
\textrm{Eq. } (\ref{rules}) \vspace{-12pt} \\
\textrm{Eq. }(\ref{rules})_1, (\ref{stiffNew1})\vspace{-12pt} \\
\textrm{Eq. }(\ref{rules})_1,(\ref{rulesPAW}) \vspace{-12pt} \\
\textrm{Eq. }(\ref{rules})_1,(\ref{StiffEnPlast})%
\end{array}
$ & \vline $\left( 1+\frac{\gamma _{s}}{\beta _{s}}\textrm{
}sgn\left(\mathbf{Z}^T\cdot\dot{\mathbf{u}}\right) \right)
\left\vert \mathbf{Z}^T\cdot\mathbf{\dot{u}}\right\vert \left\Vert \mathbf{Z}\right\Vert ^{n-2}$, \ $n>0$ \\
\hline
\end{tabular}
\caption{Summary of the 2-DoF BW models for structures. Both the
conditions on the parameter $\gamma_s$ and on the stiffness
degradation rule recalled in the caption of Table
\protect\ref{tabRiass1} hold for 2-DoF models too. The last
definition of $\dot{\xi}_s$ is named here of the KBC-type.}
\label{tabRiass1_1}
\end{table}

\clearpage

\begin{table}[tbp]
\begin{tabular}{llll}
\hline
& \vline Eq. & \vline $\nu ,$ $\eta $ & \vline $\dot{\xi}$ \\ \hline
\citep{Valanis71} & \vline (\ref{bouc type}) & \vline $1$ \ , \ $1$ & \vline
$\left\Vert dev\left( \mbox{\boldmath
$\dot{\varepsilon}$}\right) \right\Vert $ \\
$%
\begin{array}{l}
\textrm{\citep{Karray89}} \vspace{-12pt} \\
\textrm{and \citep{Casciati89}}%
\end{array}
$ & \vline (\ref{bouc type}) & \vline $1$ \ , \ $1$ & \vline $\left( 1+\frac{%
\gamma }{\beta }\textrm{ }sgn\left( \mathbf{z:}dev\left(
\mbox{\boldmath $\dot{\varepsilon}$}\right) \right) \right)
\left\vert \mathbf{z}:dev\left( \mbox{\boldmath
$\dot{\varepsilon}$}\right) \right\vert \left\Vert
\mathbf{z}\right\Vert
^{n-2} $, $n>0$ \\
\citep{ErlicherPoint2004} & \vline (\ref{bouc type}) & \vline$\textrm{Eq. } ( %
\ref{rulesTens}) $ & \vline $\left( 1+\frac{\gamma }{\beta }\textrm{
} sgn\left( \mathbf{z:}dev\left( \mbox{\boldmath
$\dot{\varepsilon}$}\right) \right) \right) \left\vert \mathbf{z}:dev\left( %
\mbox{\boldmath $\dot{\varepsilon}$}\right) \right\vert \left\Vert \mathbf{z}%
\right\Vert ^{n-2} $, $n>0$ \\
This paper & \vline (\ref{bouc type}) & \vline$\textrm{Eqs. }(\ref{rulesTens}%
)_1, (\ref{stiffNew}) $ & \vline $\left( 1+\frac{\gamma }{\beta }\textrm{ }%
sgn\left( \mathbf{z:}dev\left( \mbox{\boldmath $\dot{\varepsilon}$}\right)
\right) \right) \left\vert \mathbf{z}:dev\left(
\mbox{\boldmath
$\dot{\varepsilon}$}\right) \right\vert \left\Vert \mathbf{z}\right\Vert
^{n-2} $, $n>0$ \\ \hline
\end{tabular}%
\caption{Summary of the BW stress-strain models. Observe that the
first BW model coincides with the basic endochronic model proposed
by \citet{Valanis71}. $\dot{\protect\xi}\geq 0$, provided that
$\left\vert \protect\gamma \right\vert \leq \protect\beta $. For a
thermodynamically admissible model, the stiffness degradation rule
fulfills the condition $\dot{\protect\eta} \leq \protect\beta
\protect\nu \dot{\protect\xi}$ (see equation (\ref{damCond})). }
\label{tabRiass2}
\end{table}

\clearpage

\begin{table}[tbp]
\begin{tabular}{l|l|l}
\hline
Stress-strain law (\ref{bouc type}) & Force-displ. laws (\ref{1D})-(\ref{2D})
& Bearing shear device \\ \hline
$\mbox{\boldmath $\sigma$}$ $[Pa]$ & $w_{x},w_{y}$ $[N]$ & $w_{x}=\sigma
_{xz}\textrm{ }S,$ $Z_{y}=\sigma _{yz}\textrm{ }S$ \\
$\mathbf{z}$ $[Pa]$ & $Z_{x},Z_{y}$ $[N]$ & $Z_{x}=z_{xz}\textrm{ }S,$ $%
Z_{y}=z_{yz}\textrm{ }S$ \\
$\mbox{\boldmath $\varepsilon$}$ & $u_{x},u_{y}[m]$ & $u_{x}=2h\varepsilon
_{xz},$ $u_{y}=2h\varepsilon _{yz}$ \\
$\mathbf{E}$ $(E_{1},E_{2})$ $[Pa]$ & $A_{0}$ $[N/m]$ & $A_{0}=E_{2}S/h$ \\
$\mathbf{C}$ $(K,G)$ $[Pa]$ & $A$ $[N/m]$ & $A=GS/h$ \\
$\eta =\frac{1}{1-D}, \ \ \nu$ & $\eta _{s}:=\frac{1}{1-D_{s}},\ \ \nu_s$ & $%
\eta _{s}:=\eta, \ \ \nu _{s}:=\nu$ \\
$\dot{\xi}\left[ Pa^{n-1}s^{-1}\right] $ & $\dot{\xi}_{s}\left[
N^{n-1}ms^{-1}\right] $ & $\dot{\xi}_{s}:=\frac{\beta }{\beta _{s}}\dot{\xi}$
\\
$\beta \lbrack Pa^{1-n}]$ & $\beta _{s} \ [N^{1-n}m^{-1}]$ & $\beta _{s}=%
\frac{S^{1-n}}{h}\left( \sqrt{2}\right) ^{n-2}\beta $ \\
$\gamma \lbrack Pa^{1-n}]$ & $\gamma _{s} \ [N^{1-n}m^{-1}]$ & $\gamma _{s}:=%
\frac{\gamma }{\beta }\beta _{s}=\frac{S^{1-n}}{h}\left( \sqrt{2}\right)
^{n-2}\gamma $ \\
$\dot{\zeta}\left[ Pa^{n-1}s^{-1}\right] $ & $\dot{\zeta}_{s}\left[
N^{n-1}ms^{-1}\right] $ & $\dot{\zeta}_{s}:=\frac{\beta }{\beta _{s}}\dot{%
\zeta}$ \\
$e=e_{p}+e_{D}$ & $e_{s}=e_{s,p}+e_{s,D}$ & $e_{s,p}=Sh$ $e_{p},$ $%
e_{s,D}=Sh $ $e_{D}$ \\ \hline
\end{tabular}%
\caption{Stress-strain vs. force-displacement rules. In the ideal
case of a bearing device subjected to bi-directional shear, see
Figure \protect\ref{Fig_ShearDevice}b, stresses and strains are
uniform in the device volume. Both $\dot{\protect\xi}_{s}$ and
$\dot{\protect\xi}$ are postulated to be of the KBC type.
$\dot{\protect\xi}_{s},$ $\protect\eta _{s}$ and $\protect\nu _{s}$
are expressed as functions of global quantities like forces and
displacements, while $\dot{\protect\xi},$ $\protect\eta $ and
$\protect\nu $ depend on local quantities.}
\label{Table-Micro-Macro}
\end{table}

%%%%%%%%%%%%%%%%%%%%%%%%%%%%%%%%%%%%%%%%%%%%%%%%%%%%%%%%%%%%%%%%%%%%%%%%%%%%%%%%%%%%%%%%%%

\clearpage \listoffigures\clearpage

\clearpage

\clearpage

\clearpage

\clearpage

\clearpage

\clearpage

\clearpage

\clearpage

\end{document}